\documentclass[10pt,letterpaper]{article}
\usepackage[top=0.85in,left=2.75in,footskip=0.75in]{geometry}

\usepackage{amsmath,amssymb}

\usepackage{changepage}

\usepackage[utf8x]{inputenc}

\usepackage{textcomp,marvosym}

\usepackage{cite}

\usepackage{nameref,hyperref}

\usepackage[right]{lineno}

\usepackage{microtype}
\DisableLigatures[f]{encoding = *, family = * }

\usepackage[table]{xcolor}

\usepackage{array}

\usepackage{bm}
\usepackage{subcaption}
\usepackage{color}

\newcommand*\patchAmsMathEnvironmentForLineno[1]{%
	\expandafter\let\csname old#1\expandafter\endcsname\csname #1\endcsname
	\expandafter\let\csname oldend#1\expandafter\endcsname\csname end#1\endcsname
	\renewenvironment{#1}%
	{\linenomath\csname old#1\endcsname}%
	{\csname oldend#1\endcsname\endlinenomath}}% 
\newcommand*\patchBothAmsMathEnvironmentsForLineno[1]{%
	\patchAmsMathEnvironmentForLineno{#1}%
	\patchAmsMathEnvironmentForLineno{#1*}}%
\AtBeginDocument{%
	\patchBothAmsMathEnvironmentsForLineno{equation}%
	\patchBothAmsMathEnvironmentsForLineno{align}%
	\patchBothAmsMathEnvironmentsForLineno{flalign}%
	\patchBothAmsMathEnvironmentsForLineno{alignat}%
	\patchBothAmsMathEnvironmentsForLineno{gather}%
	\patchBothAmsMathEnvironmentsForLineno{multline}%
}
\newcommand\addtag{\refstepcounter{equation}\tag{\theequation}}
\allowdisplaybreaks

\newcolumntype{+}{!{\vrule width 2pt}}

\newlength\savedwidth

\usepackage{setspace} 
\raggedright
\usepackage[aboveskip=1pt,labelfont=bf,labelsep=period,justification=raggedright,singlelinecheck=off]{caption}

\makeatletter
\renewcommand{\@biblabel}[1]{\quad#1.}
\makeatother

\usepackage{lastpage,fancyhdr,graphicx}
\usepackage{epstopdf}
\fancyheadoffset[L]{2.25in}
\fancyfootoffset[L]{2.25in}
\begin{document}
\vspace*{0.2in}

\begin{flushleft}
{\Large
\textbf\newline{Recombination and mutational robustness in neutral fitness landscapes} % Please use "sentence case" for title and headings (capitalize only the first word in a title (or heading), the first word in a subtitle (or subheading), and any proper nouns).
}
\newline
\\
Alexander Klug\textsuperscript{1},
Su-Chan Park\textsuperscript{2},
Joachim Krug\textsuperscript{1*},
\\
\bigskip
\textbf{1} Institute for Biological Physics, University of Cologne, Cologne, Germany
\\
\textbf{2} Department of Physics, The Catholic University of Korea, Bucheon, Republic of Korea
\\
\bigskip

% Insert additional author notes using the symbols described below. Insert symbol callouts after author names as necessary.
% 
% Remove or comment out the author notes below if they aren't used.
%
% Primary Equal Contribution Note
%\Yinyang These authors contributed equally to this work.

% Additional Equal Contribution Note
% Also use this double-dagger symbol for special authorship notes, such as senior authorship.
%\ddag These authors also contributed equally to this work.

%\textcurrency Current Address: Dept/Program/Center, Institution Name, City, State, Country % change symbol to "\textcurrency a" if more than one current address note
% \textcurrency b Insert second current address 
% \textcurrency c Insert third current address


% Group/Consortium Author Note
%\textpilcrow Membership list can be found in the Acknowledgments section.

% Use the asterisk to denote corresponding authorship and provide email address in note below.
* jkrug@uni-koeln.de

\end{flushleft}
\section*{Abstract}
\label{ch:abstract}
Mutational robustness quantifies the effect of random mutations on
fitness. When mutational robustness is high, most mutations do not
change fitness or have only a minor effect on it. From the point of
view of fitness landscapes, robust genotypes form neutral networks of almost
equal fitness. Using deterministic population models it has been shown
that selection favors genotypes inside such networks, which results
in increased mutational robustness. Here we demonstrate that this
effect is massively enhanced by recombination. Our results are based
on a detailed analysis of mesa-shaped fitness landscapes,
where we derive precise expressions for the dependence of the
robustness on the landscape parameters for recombining and
non-recombining populations. In addition, we carry out numerical simulations
on different types of random holey landscapes as well as on an empirical fitness
landscape. We show that the mutational robustness of
a genotype generally correlates with its recombination weight, a new measure
that quantifies the likelihood for the genotype to arise from recombination.
We argue that the favorable effect of recombination on mutational
robustness is a highly universal feature that may have played an
important role in the emergence and maintenance of mechanisms of
genetic exchange. 

% Please keep the Author Summary between 150 and 200 words
% Use first person. PLOS ONE authors please skip this step. 
% Author Summary not valid for PLOS ONE submissions.   
\section*{Author summary}
\label{ch:author_summary}

Two long-standing and seemingly unrelated puzzles in evolutionary
biology concern the ubiquity of sexual reproduction and the robustness
of organisms against genetic perturbations. Using a
theoretical approach based on the concept of a fitness landscape, in
this article we argue that the two phenomena may in fact be closely
related. In our setting the hereditary information of an organism is
encoded in its genotype, which determines it to be either viable or
non-viable, and robustness is defined as the fraction of mutations
that maintain viability. Previous work has demonstrated that the purging of
non-viable genotypes from the population by natural selection leads to
a moderate increase in robustness. Here we show that genetic
recombination acting in combination with selection massively enhances
this effect, an observation that is largely independent of how
genotypes are connected by mutations. This suggests that the increase of
robustness may be a major driver underlying the evolution of sexual
recombination and other forms of genetic exchange throughout the
living world. 

% \linenumbers

% Use "Eq" instead of "Equation" for equation citations.
\section*{Introduction}
The reshuffling of genetic material by recombination is a ubiquitous
part of the evolutionary process across the entire range of organismal
complexity. Starting with viruses as the simplest evolving entities,
recombination occurs largely at random during the coinfection of a
cell by more than one virus strain \cite{Simon2011}. For bacteria the
mechanisms involved in recombination are already more elaborate and
present themselves in the form of transformation, transduction and
conjugation \cite{Redfield2001,Feil2001}. In eukaryotic organisms, sexual
reproduction is a nearly universal feature, and recombination is often
a necessary condition for the creation of offspring. 
Although its prevalence in nature is undeniable, the evolution and
maintenance of sex is surprising since compared to an asexual
population, only half of a sexual population is able to bear offspring
and additionally a suitable partner needs to be found \cite{MS1978,Michod1988}. 
Whereas the resulting two-fold cost of sex applies only to organisms
with differentiated sexes \cite{Holmes2018}, the fact that genetic
reshuffling may break up favorable genetic combinations or introduce
harmful variants into the genome poses a problem also to recombining
microbes that reproduce asexually \cite{Engelmoer2013,Moradigaravand2013}. Since this
dilemma was noticed early on in the development of evolutionary theory, many attempts
have been undertaken to identify evolutionary benefits of sex and
recombination based on general population genetic principles
\cite{Weismann1889,Muller1932,Muller1964,Felsenstein1974,Kondrashov1988,Kondrashov1993,Feldman1997,Burt2000,Otto2002,deVisser2007,Otto2009}. 

In this article we approach the evolutionary role of recombination from the
perspective of fitness landscapes. The fitness landscape is a mapping
from genotype to fitness, which encodes the
epistatic interactions between mutations and provides a succinct
representation of the possible evolutionary trajectories \cite{deVisser2014}. 
Previous computational studies addressing the effect of
recombination on populations evolving
in epistatic fitness landscapes have revealed a rather complex
picture, where evolutionary adaptation can be impeded or facilitated depending on,
e.g., the structure of the landscape, the rate of recombination or the time
frame of observation
\cite{Kondrashov2001,deVisser2009,Misevic2009,Moradigaravand2012,Moradigaravand2014,Nowak2014}. 

Here we focus specifically on the possible benefit of recombination
that derives from its ability to enhance the mutational robustness of
the population.
A living system is said to be robust if it is able to maintain its
function in the presence of perturbations\cite{deVisser2003,Kitano2004,Wagner2005,Lenski2006,Masel2010}. In the
case of mutational robustness these perturbations are genetic, and the
robustness of a genotype is quantified by the number of mutations that
it can tolerate without an appreciable change in fitness. Robust
genotypes that are connected by mutations therefore form plateaux in
the fitness landscape that are commonly referred to as neutral
networks\cite{vanNimwegen1999,Bornberg1999,Wilke2001,Gavrilets2004}. Mutational robustness is known to be abundant at various
levels of biological organization, but its origins are
not well understood. In particular, it is not clear if mutational
robustness should be viewed as an evolutionary adaptation, or rather
reflects the intrinsic structural constraints of living systems.   

Arguments in favor of an adaptive origin of robustness were presented
by van Nimwegen \textit{et al.} \cite{vanNimwegen1999} and by Bornberg-Bauer and Chan \cite{Bornberg1999}, who showed that selection tends to
concentrate populations in regions of a neutral network where
robustness is higher than average. Whereas this result is widely
appreciated, the role of recombination for the evolution of robustness
has received much less attention. 
An early contribution  that can be mentioned in this context is due to Boerlijst
\textit{et al.}\cite{Boerlijst1996},  who discuss the error threshold in a
viral quasi-species model with recombination and point out in a side
note that ``\textit{in sequence space recombination is always inwards
	pointing}.'' 
This observation was picked up by Wilke and Adami\cite{Wilke2003} in a
review on the evolution of mutational robustness, 
where they conjecture that the enhancement of robustness by selection
should be further amplified by recombination,  
because ``\textit{recombination alone always creates sequences that
	are within the boundaries of the current mutant cloud}.'' 
At about the same time, de Visser \textit{et al.} discussed a mechanism based on the spreading of robustness modifier alleles in recombining populations \cite{deVisser2003}
(see also \cite{Gardner2006}). 

In fact indications of a positive effect of recombination on robustness had been reported 
earlier in computational studies of 
the evolution of RNA secondary structure \cite{Huynen1994} and 2D lattice proteins \cite{Xia2002} 
in the presence and absence of recombination. 
In these systems the native folding structure of a given sequence is determined by its
global free energy minimum. Due to the restricted number of attainable
folds, most structures are degenerate in the sense that
many sequences fold into the same structure. These sequences form neutral
networks in sequence space. Xia and Levitt\cite{Xia2002} consider two scenarios, in
which the evolution of the lattice proteins is dominated by mutation
and by recombination, respectively. The results show that in
the latter case the concentration of thermodynamically stable protein
sequences is enhanced, which is qualitatively explained by the fact
that recombination tends to focus the sequences near the center of
their respective neutral network. Therefore most often a single mutation
does not change the folding structure.  

More recently, Azevedo \textit{et al.} \cite{Azevedo2006} used a
model of gene regulatory networks to investigate the origin of negative
epistasis, which is a requirement for the advantage of recombination
according to the mutational deterministic hypothesis
\cite{Kondrashov1988}. In this study a gene network is encoded by a matrix of
interaction coefficients. It is defined to be viable if its
dynamics converges to a stable expression pattern and non-viable
otherwise. Thus the underlying fitness landscape is again neutral. 
Based on their simulation results the authors argue that recombination
of interaction matrices reduces the recombinational load, which in
turn leads to an increase of mutational robustness and induces
negative epistasis as a byproduct. In effect, then, recombination
selects for conditions that favor its own maintenance. Other studies
along similar lines have been reviewed in \cite{Singhal2019}.
Taken together they suggest that the positive effect of recombination on robustness may be
largely independent of the precise structure of the space of
genotypes or the genotype-phenotype map. Indeed, a related scenario
has also been described in the context of computational evolution of linear genetic programs \cite{Hu2014}. 

Finally, in a numerical study that is similar to ours in
spirit, Sz\"{o}ll\H{o}si and Der\'{e}nyi considered the effect of
recombination on the mutational robustness of populations evolving on
different types of neutral fitness landscapes
\cite{Szollosi2008}. Using neutral networks that were either generated
at random or based on RNA secondary structure, they found that
recombination generally enhances mutational robustness by a
significant amount. Moreover, they showed that this observation holds
not only for infinite populations but also for finite populations, as
long as these are sufficiently polymorphic.  

The goal of this article is to explain these scattered observations in
a systematic and quantitative way. For this purpose we begin by a
detailed examination of the simplest conceivable setting consisting of
a haploid two-locus model with three viable and one lethal
genotype \cite{Gavrilets2004}. We derive explicit expressions for the robustness as a
function of the rates of mutation and recombination that demonstrate
the basic phenomenon and guide the exploration of more complex
situations. The two-locus results are then generalized to mesa
landscapes with $L$ diallelic loci, where genotypes carrying up to $k$
mutations are viable and of equal fitness
\cite{Gerland2002,Peliti2002,Berg2004,Wolff2009}. Using a communal recombination scheme and previous results for multilocus mutation-selection models, we arrive at precise asymptotic results
for the mutational robustness for large $L$ and small mutation rates. 
Subsequently two
types of random holey landscape models are considered, including a
novel class of sea-cliff landscapes in which the fraction of viable
genotypes depends on the distance to a reference sequence. For the isotropic percolation landscape 
analytic upper and lower bounds on the robustness are derived.

As a first step towards a unified explanation for 
the effect of recombination on mutational robustness we introduce
the concept of the recombination weight,
which is a measure for the likelihood of a genotype to arise from a
recombination event. In analogy to the classic fitness landscape concept in the context of selection \cite{deVisser2014}, the recombination weight allows one to identify 
genotypes that are favored by recombination without referring to any specific evolutionary dynamics. 
We show that recombination weight correlates with mutational robustness for the landscape structures used in this work, thus providing a mechanistic basis
for the enhancement of robustness by recombination. 
Finally, using an empirical fitness landscape as an example, we
quantify the competition between selection and recombination as a function of
recombination rate.  
Throughout we describe
the evolutionary dynamics by a deterministic, discrete time model that 
will be introduced in the next section.

\section*{Models and Methods}

\subsection*{Genotype space}
We consider a haploid genome with $L$ loci and the corresponding genotype is represented by a sequence
$\sigma=(\sigma_1, \sigma_2,...,\sigma_L)$
of length $L$. The index $i$ labels genetic loci and each locus
carries an allele specified by $\sigma_i$. 
Here we rely on binary sequences, which means that there are only two
different alleles $\sigma_i \in \{0, 1\}$. This can be either seen as
a simplification in the sense that only two alleles are assumed to
exist, or in the sense that the genome consisting of all zeros
describes the wild type, and the 1's in the sequence display mutations for which no further distinctions are made. 

The resulting genotype space is a hypercube of dimension $L$, 
where the $2^L$ genotypes represent vertices, and two genotypes that
differ at a single locus and are mutually reachable by a point
mutation are connected by an edge. 
A metric is introduced by the Hamming distance 
\begin{equation}
	d(\sigma,\kappa)=\sum_{i}(1-\delta_{\sigma_i  \kappa_i}),
\end{equation}
which measures the number of point mutations that separate two
genotypes $\sigma$ and $\kappa$.
Here and in the following the Kronecker symbol is defined as
$\delta_{xy} = 1$ if $x = y$ and $\delta_{xy} = 0$ otherwise.
The genotype $\bar{\sigma}$ at maximal distance $d(\sigma,\bar{\sigma}) = L$ from a given genotype
$\sigma$ is called its antipodal, and can be defined by
$\bar{\sigma}_i = 1 - \sigma_i$.  
Finally, in order to generate a fitness landscape, a (Wrightian)
fitness value $w_\sigma$ is assigned to each genotype.

\subsection*{Dynamics}
The forces that drive evolution are selection, mutation and recombination. 
To model the dynamics we use a deterministic, discrete-time model with non-overlapping generations,  
which can be viewed as an infinite population limit of the Wright-Fisher model. 
% of Wright-Fisher type.
Demographic stochasticity or genetic drift is thus neglected. Numerical simulations of evolution on
neutral networks have shown that the infinite population dynamics
is already observable for moderate population sizes, which justifies
this approximation \cite{vanNimwegen1999,Szollosi2008}. We will return to this point in the Discussion. 

Once the frequency $f_\sigma(t)$ of a genotype $\sigma$ at generation
$t$ is given, the frequency at the next generation
is determined in three steps representing selection, mutation, and recombination.
After the selection step, the frequency $q_\sigma(t)$ is given as
\begin{align}
	q_\sigma(t) = \frac{w_\sigma}{\bar w(t)} f_\sigma(t),
\end{align}
where $\bar w \equiv \sum_\sigma w_\sigma f_\sigma(t)$ is the mean
population fitness at generation $t$.
After the mutation step, the frequency $p_\sigma(t)$ is given as
\begin{align}
	p_\sigma(t) = \sum_{\kappa} U_{\sigma\kappa} q_\kappa(t),
\end{align}
where $U_{\sigma\kappa}$ is the probability that an individual with genotype $\kappa$ mutates to genotype $\sigma$ in one generation. 
Here, we assume that alleles at each locus mutate independently, and
the mutation probability $\mu$
is the same in both directions ($0 \to 1$ and $1 \to 0$) and across loci. This leads to the
symmetric mutation matrix
\begin{equation}
	U_{\sigma \kappa} = (1-\mu)^{L-d(\sigma, \kappa)} \mu^{d(\sigma, \kappa)}.
	\label{Eq:mut_scheme}
\end{equation}
\begin{figure}
	\centering
	\includegraphics[width=0.7\linewidth]{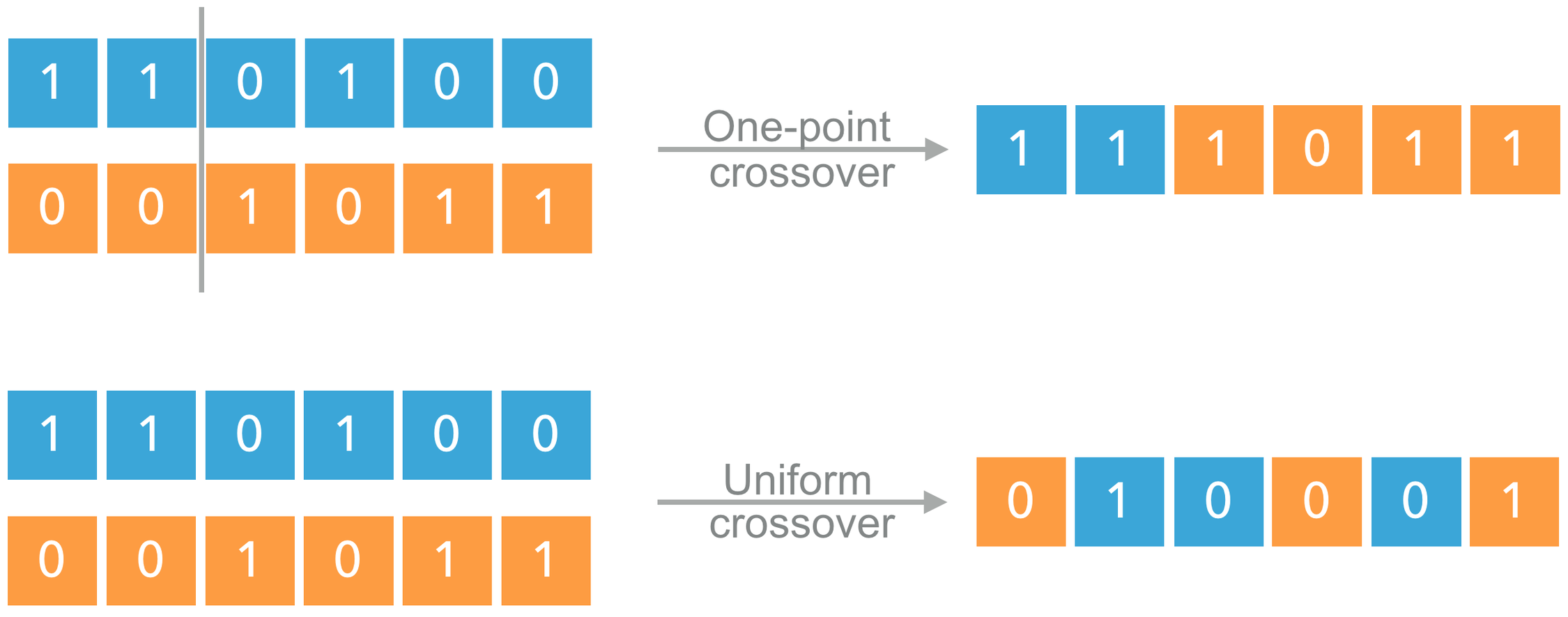}
	%\captionsetup{justification=centering}
	\caption[Source1]{\textbf{Recombination
			schemes.} 
		In the one-point crossover scheme, the parent genotypes are cut
		once between two randomly chosen loci and recombined to form the offspring. In the uniform crossover scheme, at each locus of the offspring, an allele present in one of the parents is chosen at random.}
	\label{fig:recscheme}
\end{figure}
In order to incorporate recombination we have to consider the
probability that two parents with respective genotypes $\kappa$ and
$\tau$ beget a progeny with genotype $\sigma$ by recombination. 
This is represented by the following equation:
\begin{equation}
	\label{eq:RecTensor}  
	f_{\sigma}(t+1)=\sum_{\kappa \tau} R_{\sigma|\kappa \tau} p_{\kappa}(t) p_{\tau}(t) .
\end{equation}
Descriptively speaking, two genotypes ($\kappa$ and $\tau$) are taken to recombine with a probability that is equal to their frequency in the population (after selection and mutation). The probability for the offspring genotype $\sigma$ is then given by  $R_{\sigma|\kappa \tau}$.
These probabilities depend of course on the parent genotypes $\kappa$ and $\tau$ but also on the recombination scheme. Here we consider a uniform and a one-point crossover scheme;
see Fig~\ref{fig:recscheme} for a graphical representation. 
These two represent extremes in a spectrum of possible recombination
schemes. Nevertheless we will show that both lead to qualitatively similar 
results in the regimes of interest. In the case of uniform crossover
the recombination probabilities are given by
\begin{equation} \label{eq:uniformcrossover2}
	R_{\sigma|\kappa \tau}=\frac{r}{2^L} \left(\prod_i^L  (\delta_{\sigma_i \kappa_i}+\delta_{\sigma_i \tau_i}) \right)+\frac{1-r}{2}\left(\delta_{\sigma \kappa} +\delta_{\sigma \tau}\right) 
\end{equation}
and in the case of one point crossover the probabilities can be written as
\begin{align} \label{eq:onepointcrossover}
	\begin{aligned}
		R_{\sigma|\kappa \tau}={} & \frac{r}{2(L-1)}\sum_{n=1}^{L-1} \Biggl[\left(\prod_{m=1}^{n} \delta_{\sigma_{m} \kappa_{m}}  \right) \left( \prod_{m=n+1}^{L} \delta_{\sigma_{m} \tau_{m }}    \right) \\
		+&\left(\prod_{m=1}^{n} \delta_{\sigma_{m} \tau_{m}}  \right) \left( \prod_{m=n+1}^{L} \delta_{\sigma_{m} \kappa_{m}}    \right)  \Biggr] + \quad\frac{1-r}{2}\left(\delta_{\sigma \kappa} +\delta_{\sigma \tau}\right).  \\
	\end{aligned}
\end{align}
In both equations a variable $r \in \left[0, 1\right]$ appears which describes the recombination rate. For $r=0$ no recombination occurs and $f_\sigma(t+1)$ is the same as $p_\sigma(t)$.
For $r=1$ recombination is a necessary condition for the creation of
offspring (obligate recombination). 
But also intermediate values of $r$ can be chosen as they occur in nature, e.g., for bacteria and viruses. 

In the following we are mostly interested in the equilibrium frequency
distribution $f_\sigma^*$ of a population,
which is determined by the stationarity condition 
\begin{equation}
	\label{eq:stationarity}
	f_\sigma(t+1)=f_\sigma(t)=f_\sigma^* 
\end{equation}
for all genotypes $\sigma$.

\subsection*{Mutational robustness}
From the point of view of fitness landscapes the occurrence of
mutational robustness implies that fitness values of neighboring
genotypes are degenerate, giving rise to neutral networks in genotype
space \cite{Wagner2005,vanNimwegen1999,Bornberg1999,Wilke2001,Gavrilets2004}. In
order to model this situation we use two-level landscapes that only
differentiate between genotypes that are viable ($w_\sigma = 1$) or
lethal ($w_\sigma = 0$). Any selective advantage between viable
genotypes is assumed to be negligible. The mutational robustness of a
population can then be measured by the average fraction of viable point
mutations in an individual, which depends on the population
distribution in genotype space \cite{vanNimwegen1999,Bornberg1999,Wilke2001}. It increases if the population mainly adapts to genotypes for which most point mutations are viable. 
Therefore we define mutational robustness $m$ as the average fraction of viable point mutations of a population,
\begin{equation}
m\equiv \sum_{\sigma \in V}m_\sigma f_\sigma^* \quad \textrm{with}  \quad
m_\sigma \equiv \frac{n_\sigma}{L}. 
	\label{Eq:m}
\end{equation}	
Here the sum is over the set $V$ of all viable genotypes and
$n_{\sigma}$ is the number of
viable point mutations of genotype $\sigma$. We will 
refer to $m_\sigma$ as the mutational robustness of the genotype. The expression is normalized
by the total number of loci $L$, since in an optimal setting the
entire population has $L$ viable genotypic neighbors and $m_\sigma =
1$ for all $\sigma \in V$. The value of
$m$ is thus constrained to be in the range $[0, 1]$. We weight the
genotypes by their stationary frequencies $f_\sigma^\ast$, since we want to determine
the mutational robustness of populations that are in equilibrium with their environment.

\subsection*{Recombination weight} 
In order to elucidate the interplay of recombination and mutational
robustness it will prove helpful to introduce a representation of how
recombination can transfer genotypes into each other.
The number of distinct genotypes that two recombining genotypes are
able to create depends on their Hamming distance. In particular, the
recombination of two identical genotypes does not create any novelty,
whereas a genotype and its antipodal are able
to generate all possible genotypes through uniform crossover.

Here we introduce a measure which expresses how many pairs of viable
genotypes are able to recombine to a specific genotype. It is
complementary to the mutational robustness, in the sense that instead
of counting the viable mutation neighbors of a genotype, the size of its
recombinational neighborhood of viable recombination pairs is
determined. The recombinational neighborhood depends on the
recombination scheme and the distribution of viable genotypes in the
genotype space.
For a given recombination scheme the probability for a genotype
$\sigma $ to be the outcome of recombination of two genotypes $\kappa,
\tau$ is given by the recombination tensor $R_{\sigma|\kappa \tau}$.
The \textit{recombination weight} $\lambda_{\sigma}$ is therefore
obtained by summing the recombination tensor over all ordered pairs of
viable genotypes,
\begin{equation}
	\label{eq:recombweight} 
	\lambda_{\sigma}=\frac{1}{2^L}\sum_{\kappa \in V, \tau \in V} R_{\sigma|\kappa \tau}.
\end{equation}
It can be seen from \eqref{eq:RecTensor} that $\lambda_\sigma = 1$ when all
genotypes are viable, and hence the normalization by $2^{L}$ ensures
that the recombination weight lies in the range $[0,1]$.
Under this normalization, the recombination weights sum to
$\sum_\sigma \lambda_\sigma = \vert V \vert^2/2^L$, where $\vert V \vert$ stands for
the number of viable genotypes.
In the following the genotype maximizing $\lambda_\sigma$ will be
referred to as the \textit{recombination center} of the landscape. 

Since neutral landscapes only differentiate between viable (unit fitness) and
lethal (zero fitness) genotypes, 
the recombination weight
\eqref{eq:recombweight} can alternatively be written as a sum over all ordered
pairs of genotypes whereby the recombination tensor is multiplied by the
pair's respective fitness,
\begin{equation}
	\label{eq:recombweight2} 
	\lambda_{\sigma}=\frac{1}{2^L}\sum_{\kappa,  \tau} R_{\sigma|\kappa \tau} w_{\kappa} w_{\tau}.
\end{equation}
In this way the concept naturally
generalizes to arbitrary fitness landscapes. 
In the absence of
recombination ($r=0$) the recombination weight \eqref{eq:recombweight2} of a genotype is simply proportional
to its fitness, $\lambda_\sigma = \tilde{w} w_\sigma$, where
$\tilde{w} = 2^{-L} \sum_\sigma w_\sigma$ is the unweighted average fitness.
Within our recombination schemes, the recombination tensor depends linearly
on $r$ and, by definition, so does the recombination weight. 
Accordingly, for general $r$ the recombination weight interpolates
linearly between the limiting values at $r=0$ and $r=1$. 
Since $\lambda_\sigma$ for $r=0$ is known, the remaining task
will be to find $\lambda_\sigma$ for $r=1$.

\section*{Results}
In the following sections we investigate how mutational robustness 
depends on the mutation and recombination rates. 
In order to test the generality of our results, we use, besides
contrasting recombination schemes, also different neutral landscape
models such as the mesa
\cite{Gerland2002,Peliti2002,Berg2004,Wolff2009} and the percolation models \cite{Gavrilets2004,Gavrilets1997}. Additionally we introduce a more general landscape
named sea-cliff model, which combines elements of both the landscape models and contains them as limiting cases.
In the end, we discuss mutational robustness and its relation with recombination weight for an empirical landscape.

Two-locus models are commonly used in population genetics to gain a
foothold in understanding evolutionary scenarios involving multiple recombining
loci~\cite{Gavrilets2004,Gardner2006,Buerger2000,Crow1965,Eshel1970,Kimura1985,Higgs1998,Park2011,Altland2011,Weinreich2013}. 
Following this tradition, we first discuss a two-locus model and then extend our results to multi-locus models.

\subsection*{Two-locus model}
\begin{figure}
	\centering
	\includegraphics[width=0.35\linewidth]{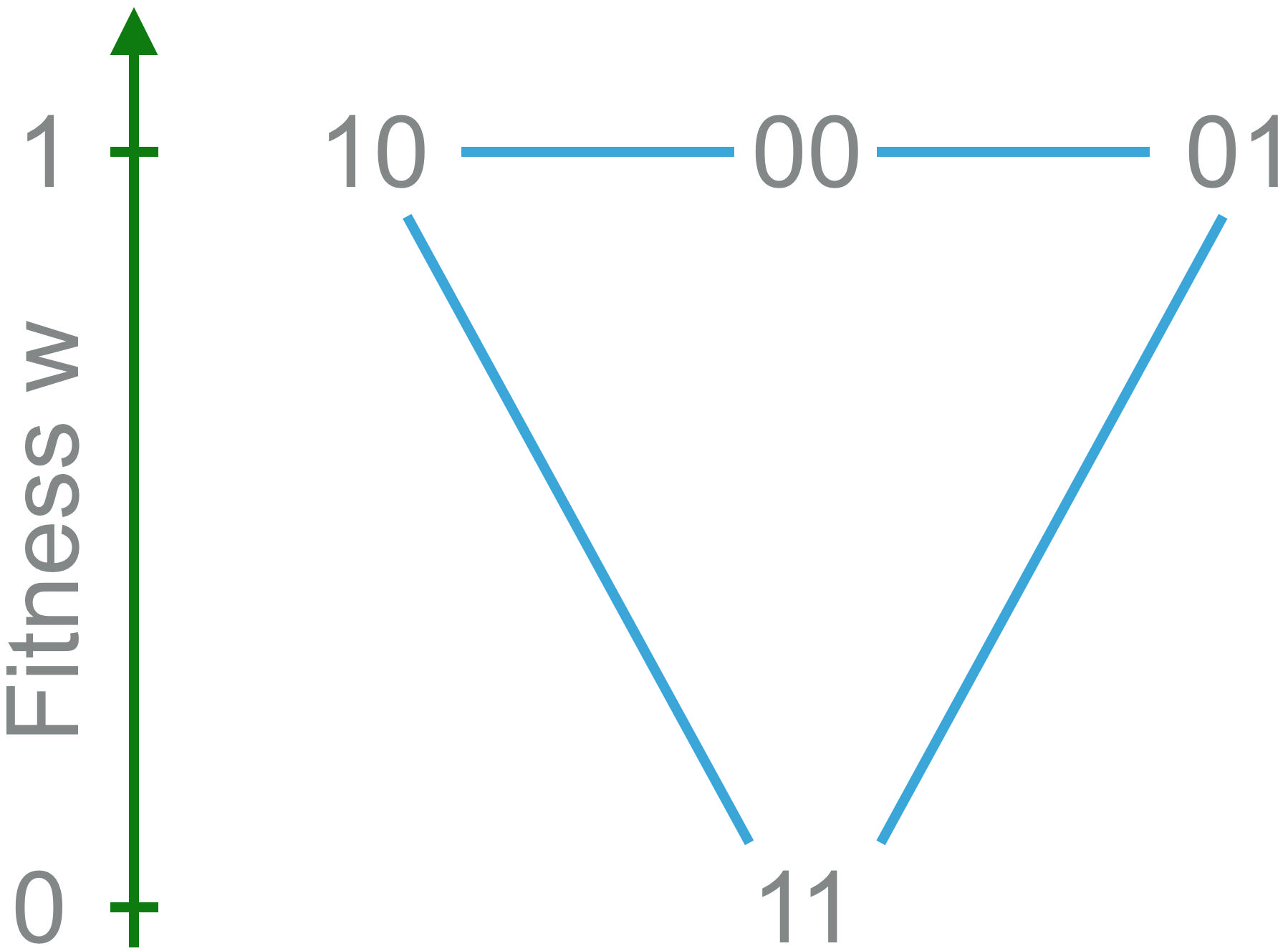}
	\caption[Source2]{\textbf{Two-locus
			model.} Genotype (1,1) is lethal while the other three
		genotypes are viable with the same fitness. Here,
		genotype (0,0) is most robust since both its single mutants are viable.}
	\label{fig:twolocusmodel}
\end{figure}

The simplest fitness landscape to study the mutational robustness of a
population would be the haploid two-locus model in which all but one
genotype are viable \cite{Gavrilets2004}; see
Fig~\ref{fig:twolocusmodel} for a graphical representation of the model. 
In this setting the population gains mutational robustness if the frequency of 
the genotype (0,0) for which both point mutations are viable increases
relative to the genotypes (0,1) and (1,0). This model has been analyzed
previously using a unidirectional mutation scheme where
reversions $1 \to 0$ are suppressed \cite{Nowak1997,Phillips1998}. As a
consequence, selection cannot contribute to mutational robustness
because the genotype (0,0) goes extinct in the absence of
recombination. Here we consider the case of bidirectional, symmetric mutations
in which both selection and recombination contribute to robustness. 
A comparison of the two mutation schemes is provided in S1~Appendix. 

We proceed to solve the equilibrium condition Eq~\eqref{eq:stationarity}. 
Since the equilibrium genotype frequencies $f_{01}^*$ and $f_{10}^*$
are the same due to the symmetry of the landscape and the mutation
scheme, the recombination step at stationarity reads 
\begin{align*}
	f_{00}^*= p_{00} - \rho (p_{00}p_{11} - p_{10}p_{01}) \quad &\Leftrightarrow \quad f_0 = p_0 - \rho D,\\
	f_{10/01}^*= p_{10/01} + \rho (p_{00}p_{11} - p_{10}p_{01})\quad &\Leftrightarrow \quad f_1 = p_1 + 2\rho D,\\
	f_{11}^*= p_{11} - \rho (p_{00}p_{11} - p_{10}p_{01}) \quad &\Leftrightarrow \quad  f_2 = p_2 - \rho D,
	\addtag
\end{align*}
where $p_\sigma$ is the (equilibrium) frequency of genotype $\sigma$ after the mutation step,  
$f_i$ and $p_i$ are the corresponding lumped frequencies
\cite{Baake2007} of all genotypes with $i$ $1$'s,
and $D \equiv p_{00}p_{11} - p_{10}p_{01}=p_0 p_2 - p_1^2/4$ is the
linkage disequilibrium after the mutation step. 
Notice that the one-point and uniform crossover schemes give the same equation form except that the parameter $\rho$
is given by $\rho=r$ in the case of one-point crossover and $\rho = r/2$ for
uniform crossover. However, we would like to emphasize that this is a mere coincidence of the two-locus model which disappears as soon as $L$ is larger than 2.

The lumped frequencies $q_i$ of all genotypes with $i$ 1's after the selection step 
are given by
\begin{align}
	q_0 = \frac{f_0}{1 - f_2}, \quad q_1 = \frac{f_1}{1-f_2}, \quad q_2 = 0.
\end{align}
Applying the mutation step we obtain
\begin{align*}
	p_0 &= q_0 {( 1 -  \mu)}^2 + \mu(1-\mu) q_1 
	= \mu(1-\mu) + (1-\mu)(1-2\mu) q_0,\\
	p_1 &= q_1 \left [ {( 1 - \mu)}^2 + \mu^2 \right ] + 2 \mu(1-\mu) q_0 
	= 1 - 2\mu +2 \mu^2 - {(1 - 2 \mu)}^2 q_0,\\
	p_2 &= \mu (1-\mu) q_1 + \mu^2 q_0 = \mu(1-\mu) - \mu (1 - 2\mu) q_0,\\
	D&=p_0 p_2 - p_1^2/4 = - \frac{1}{4} {(1 - 2\mu)}^2 {(1-q_0)}^2,
	\addtag
\end{align*}
where we have used the normalization $q_0 + q_1 = 1$ to express the right hand sides in terms of 
$q_0$. Putting everything together, the problem is reduced to 
solving the following third order polynomial equation for $q_0$,
\begin{align}
	0 &= q_0 (1-f_2) - f_0 = q_0 ( 1 - p_2 + \rho D) - p_0 + \rho D\nonumber \\
	&=\frac{\rho}{4} {(1 - 2\mu)}^2 {( 1 - q_0)}^2 (1 + q_0)
	+\mu \left [ 1 - 2 q_0 - q_0^2 - \mu (1-2 q_0) (1+q_0) \right ],
	\label{Eq:two_locus_q0}
\end{align}
from which we can in principle find exact analytic expressions for $f_\sigma^*$. However, it is difficult to extract useful information from the exact solution. 
In the following we will therefore provide approximate solutions.

If we neglect recombination ($\rho=0$), we obtain the following equilibrium genotype frequency distribution:
\begin{align*}
	f_{00}^*(\rho=0) &= 
	\frac{1-\mu}{2}\sqrt{8-16\mu + 9 \mu^2} - \frac{1}{2} \left ( 2 - 5 \mu + 3 \mu^2\right )\\
	&\approx \left(\sqrt{2}-1\right)+\left(\frac{5}{2}-2 \sqrt{2}\right) \mu +O\left(\mu ^2\right), \\
	f_{01/10}^*(\rho=0)&= 
	\frac{1}{4} \left ( 4 - 9 \mu + 6\mu^2 \right ) - \frac{1 - 2\mu}{4} \sqrt{8-16\mu + 9 \mu^2} \\
	&\approx \left(1-\frac{1}{\sqrt{2}}\right)+\left(\frac{3}{\sqrt{2}}-\frac{9}{4}\right) \mu +O\left(\mu ^2\right),\\
	f_{11}^*(\rho=0) &= \frac{\mu}{2} \left(4-3 \mu-\sqrt{8-16 \mu +9 \mu ^2} \right) 
	\approx \left(2-\sqrt{2}\right) \mu +O\left(\mu ^2\right).
	\addtag
\end{align*}

\begin{figure}%[!h]
	\centering
	\includegraphics[width=1.0\linewidth]{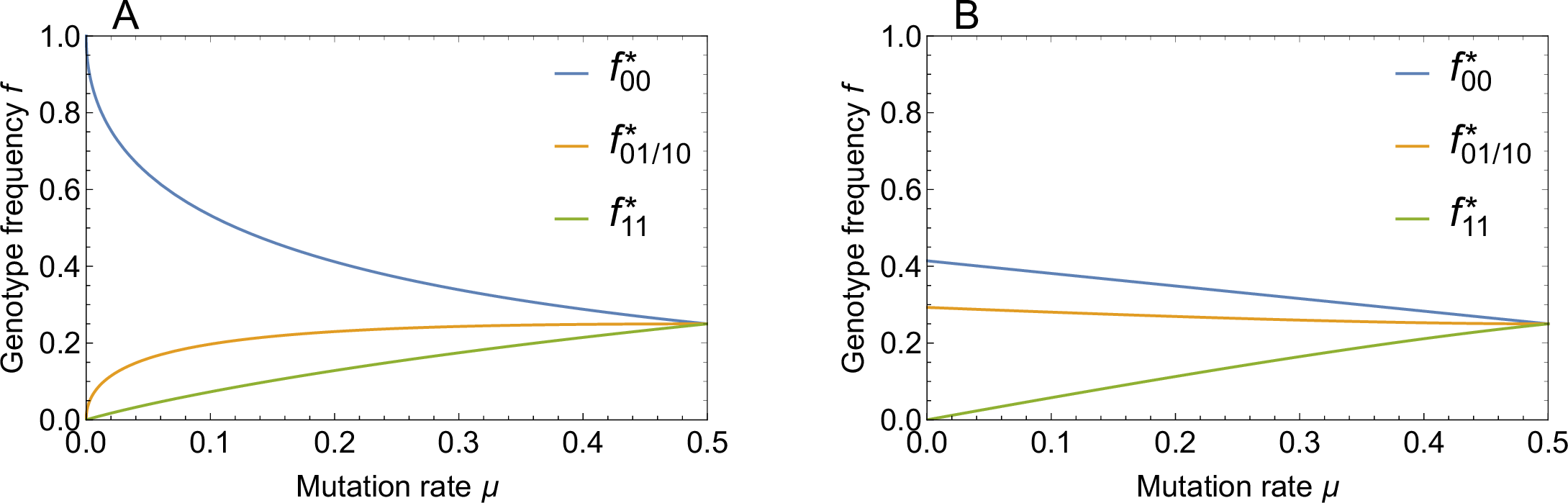}
	\caption[Source1]{\textbf{Equilibrium genotype frequencies in
			the two locus model.} Genotype frequencies in the
		stationary state are shown as a function of mutation rate for
		(A) strong recombination ($\rho=1$) and (B) no
		recombination ($\rho=0$).}
	\label{fig:2lfrequencies}
\end{figure}

When $\rho=1$, which corresponds to the one-point crossover scheme with $r=1$, 
linkage equilibrium ($f_{00}f_{11} = f_{10} f_{01}$) is restored after one generation~\cite{Park2011}.
Accordingly, we can treat  each locus independently and get rather simple expressions for $f^*_\sigma$ as
\begin{align*}
	\label{Eq:frequencies_rho1}
	f_{00}^*(\rho=1)&=\frac{1}{4} {\left(2+\mu -\sqrt{\mu^2 +4\mu}\right)}^2 \approx 1-2 \sqrt{\mu }+2 \mu +O\left(\mu ^{3/2}\right),\\
	f_{01/10}^*(\rho=1)&=\frac{1}{4} \left(2+\mu -\sqrt{\mu^2 +4\mu}\right) \left(\sqrt{\mu^2 +4\mu}-\mu \right) \approx \sqrt{\mu }-\frac{3 \mu }{2}+O\left(\mu ^{3/2}\right) ,\\
	f_{11}^*(\rho=1)&=\frac{1}{4} {\left(\sqrt{\mu^2 +4\mu}-\mu \right)}^2 \approx \mu -O\left(\mu^{3/2}\right).
	\addtag
\end{align*}
We depict the equilibrium solutions for the above two cases 
in Fig~\ref{fig:2lfrequencies}.

\begin{figure}%[!h]
	\centering
	\includegraphics[width=0.5\linewidth]{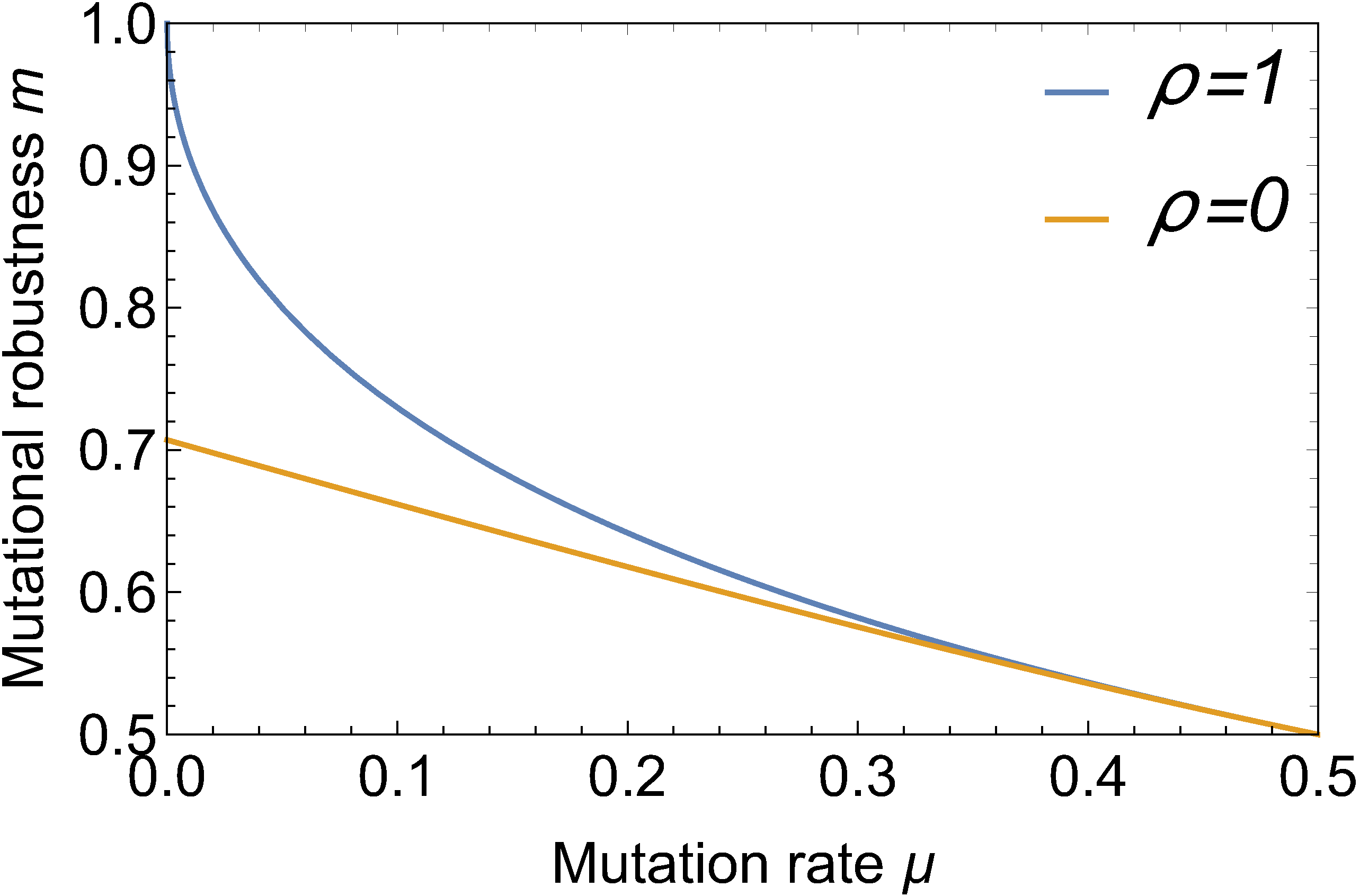}
	\caption[Source1]{\textbf{Mutational robustness as a function
			of mutation rate.} The figure shows the robustness
		in the two-locus model at $\rho = 0$ and $\rho = 1$. Recombination leads to a massive enhancement
		of robustness for small mutation rates.}
	\label{fig:mutrate}
\end{figure}

Now, the mutational robustness
\begin{equation}
	m= \frac{1}{2}(2 f_{00}^*+f_{10}^*+f_{01}^*)=f_{00}^*
	+ f_{10}^* = f_0 + \frac{1}{2} f_1
\end{equation}
for the above two cases is obtained as
\begin{align}
	m(\mu, \rho=0)&=\frac{1}{4} \left(\mu +\sqrt{8-16\mu+9 \mu^2 }\right) \approx \frac{1}{\sqrt{2}}-\left(\frac{1}{\sqrt{2}}-\frac{1}{4}\right) \mu +O\left(\mu ^2\right),\label{Eq:mr0}\\
	m(\mu, \rho=1)&=\frac{1}{2} \left(2+\mu -\sqrt{\mu^2 +4\mu}\right) \approx 1-\sqrt{\mu }+\frac{\mu }{2}+O\left(\mu ^{3/2}\right),\label{Eq:mr1}
\end{align}
which is depicted in Fig~\ref{fig:mutrate}. These results encapsulate in a simple form the main topic of this paper. Selection alone 
($\rho = 0$) leads to a moderate increase of robustness from the baseline value $m=\frac{1}{2}$ corresponding to a random 
distribution over genotypes, which is attained at $\mu = \frac{1}{2}$, to $m = \frac{1}{\sqrt{2}}$ for $\mu \to 0$. In contrast,
for recombining populations ($\rho = 1$) robustness is massively enhanced at small mutation rates 
due to the strong frequency increase of the most robust genotype (0,0) and reaches the maximal value $m=1$ 
at $\mu = 0$. The underlying mechanism is analogous to Kondrashov's deterministic mutation hypothesis, 
which posits that recombination makes selection against deleterious mutations more effective when these interact synergistically 
\cite{Kondrashov1988}. 
In the present case recombination increases the frequency of the double mutant genotype $(1,1)$, which is subsequently 
purged by selection, and thereby effectively drives the frequency
of the allele $1$ at both loci to zero.
The enhancement of the frequency of the genotype (0,0) by
recombination is also reflected in the recombination weights, which
take on the values
\begin{equation}
	\label{eq:2locus_weights}
	\lambda_{00} = \frac{3}{4} + \frac{\rho}{4}, \;\;\; \lambda_{01} =
	\lambda_{10} = \frac{3}{4} - \frac{\rho}{4}, \;\;\; \lambda_{11} =
	\frac{\rho}{4}. 
\end{equation}
Thus the genotype (0,0) is the recombination center of the two-locus landscape.

Next we investigate how mutational robustness varies with $\mu$ for intermediate recombination rates, 
assuming that $\mu$ is small. 
As can be seen from Eq~\eqref{Eq:two_locus_q0}, the asymptotic behavior of the solution for small $\rho$ and $\mu$ depends on
which of the two parameters is smaller.
We first consider the case $\rho \ll \mu \ll 1$. Defining $l = \rho/(4\mu) \ll 1$, Eq~\eqref{Eq:two_locus_q0}
is approximated by
\begin{align}
	0= 
	l{(1-q_0)}^2(1+q_0) +   1 - 2 q_0 - q_0^2  - \mu (1-2q_0)(1+q_0),
	\label{Eq:q0_rsmall}
\end{align}
where we kept terms up to $O(\mu)$, since we have not determined whether $l$ is smaller than $\mu$ or not.
Since $q_0 = \sqrt{2}-1$ is the solution of Eq~\eqref{Eq:q0_rsmall} for $l=\mu = 0$, 
we set $q_0 = \sqrt{2}-1 + a l + b \mu$ and solve the equation to leading order, which gives
\begin{equation}
	q_0 \approx \sqrt{2} - 1 + \left ( 3 - 2 \sqrt{2} \right ) l - \left ( \frac{3}{2} - \sqrt{2} \right ) \mu.
\end{equation}
The mutational robustness then follows as
\begin{equation}
	m = f_0 + \frac{f_1}{2} = \frac{1}{2} + \frac{p_0 - p_2}{2}
	=\frac{1}{2} + (1 - 2 \mu) \frac{q_0}{2}
	\approx \frac{1}{\sqrt{2}} + \frac{3-2\sqrt{2}}{2} l -\left(\frac{1}{\sqrt{2}}-\frac{1}{4}\right) \mu,
\end{equation}
which is consistent with our previous result for $\rho=0$; see Eq~\eqref{Eq:mr0}. 
We note that in this regime it is sufficient for the recombination rate to be of order $O(\mu^2)$ to compensate the negative effect of 
mutations on mutational robustness, as the two effects cancel when $\rho = \rho_c$ with
\begin{equation}
	\rho_c = 2(5+4\sqrt{2})\mu^2 \approx 21.3 \times \mu^2. 
\end{equation}

In the regime $\rho \gg \mu$, Eq~\eqref{Eq:two_locus_q0} is
approximated as
\begin{equation}
	( 1 - 4 \mu) {(1-q_0)}^2 ( 1 + q_0) + s ( 1 - 2 q_0 -q_0^2)=0,
	\label{Eq:q0_msmall}
\end{equation}
with $s=4\mu/\rho$. Again we have kept terms up to $O(\mu)$ because $\mu$ and  $s$ are of the same
order if $\rho = O(1)$. Since the solution of Eq~\eqref{Eq:q0_msmall} for $\mu = s=0$ is $q_0=1$, we set 
$q_0 = 1 - \alpha$ with $\alpha \ll 1$. Inserting this into Eq~\eqref{Eq:q0_msmall}, we get 
$\alpha \approx \sqrt{s}$.
Since $\alpha \gg \mu$, $q_0 = 1 - \sqrt{s}$ is the approximate solution to leading order. Hence
\begin{equation}
	\label{Eq:q0_msmall2}
	m = \frac{1}{2} + (1-2\mu) \frac{q_0}{2} \approx 1 - \sqrt{\frac{s}{4}} = 1  - \sqrt{\frac{\mu}{\rho}},
\end{equation}
which is again consistent with our previous result for $\rho=1$ in Eq~\eqref{Eq:mr1}. The square root dependence on $\mu/\rho$ 
derives from the corresponding behavior of the genotype frequency $f^\ast_{00}$ and has been noticed previously in the 
model with unidirectional mutations \cite{Nowak1997,Phillips1998}. 

\begin{figure}%[!h]
	\centering
	\includegraphics[width=0.8\linewidth]{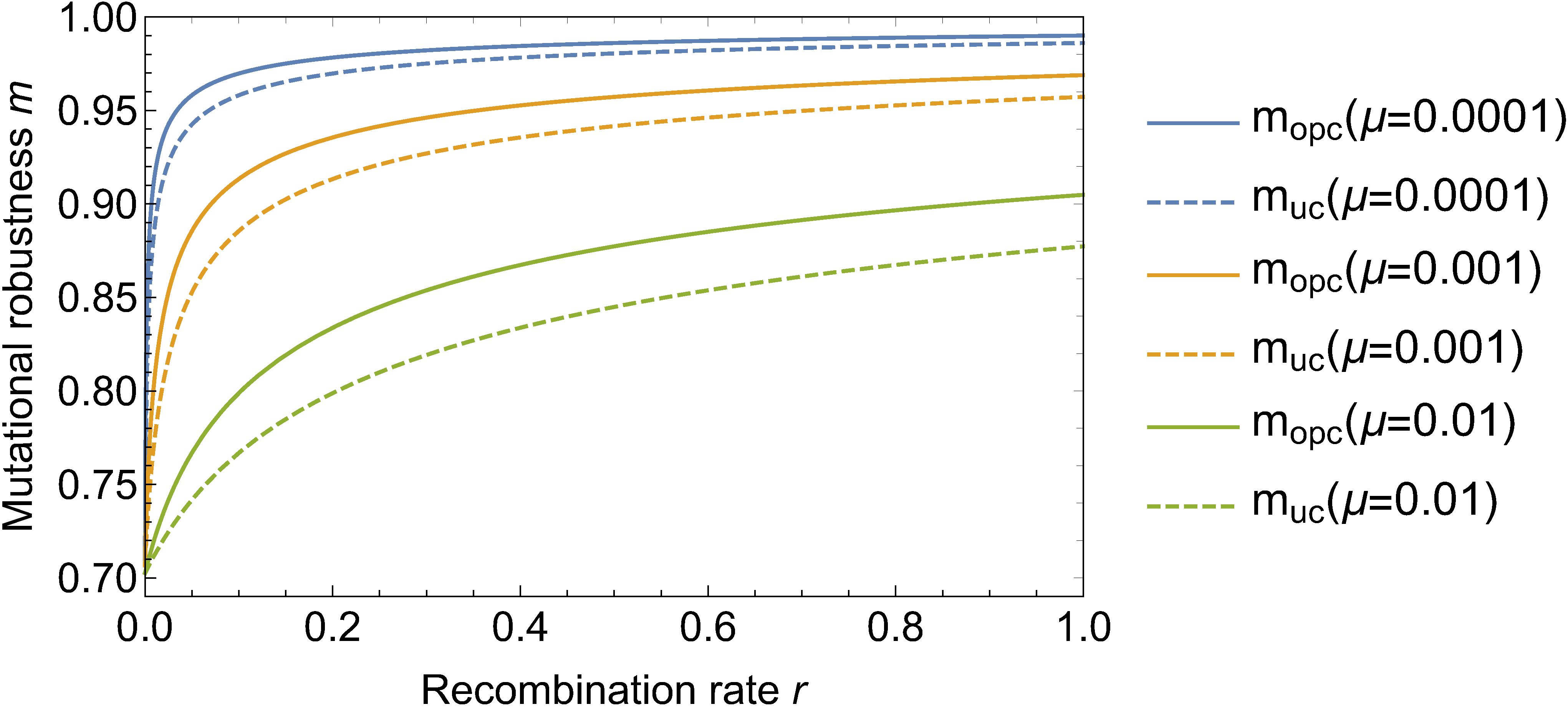}
	\caption[Source1]{\textbf{Mutational robustness as a function
			of recombination rate.}
		The figure shows the mutational robustness for one-point
		crossover ($m_\textrm{opc}$) and uniform crossover ($m_\textrm{uc}$) and
		three different values of the mutation rate $\mu$. When
		mutations are rare, a small amount of recombination is
		sufficient to significantly increase mutational robustness.}
	\label{fig:lfullrange}
\end{figure}

For arbitrary $\rho$ and $\mu$, we have to use the full Eq~\eqref{Eq:two_locus_q0}. Fig~\ref{fig:lfullrange} illustrates the behaviour of mutational robustness as a function of the recombination rate for different mutation rates and both recombination schemes. For small
$\mu$, a low rate of recombination suffices to bring the robustness close to its maximal value $m=1$. More precisely, according to
Eq~\eqref{Eq:q0_msmall2}, a robustness $m > 1 - \epsilon$ is reached for recombination rates $\rho > \mu/\epsilon^2$.

To summarize, we have seen that analytic results for the two-locus
model are easily attainable.
For multi-locus models it is much more challenging to derive analytical results, particularly in the presence of recombination.
By way of contrast the dynamics induced only by mutation and selection are easier to understand: 
While mutations increase the genotype diversity in the population, fitter ones grow in frequency through selection, which 
reduces diversity.
Although one might expect that recombination would increase diversity, a number of studies have shown 
that recombination is more likely to impede the divergence of populations. 
Recombining populations tend to cluster on single genotypes or in a limited region of a genotype space and 
furthermore the waiting times for peak shifts in multipeaked fitness landscapes 
diverge at a critical recombination rate \cite{deVisser2009,Nowak2014,Paixao2015,Higgs1998,Park2011,Altland2011}.
The results for the two-locus model presented above are consistent with this behaviour, 
as the genotype heterogeneity of the population decreases with
increasing recombination rate (S1~Fig).

In the following we will investigate how the focusing effect of recombination enhances the mutational robustness of the 
population in three different multi-locus models. 

\subsection*{Mesa landscape}
In the mesa landscape it is assumed that up to a certain number $k$ of
mutations all genotypes are functional and have unit
fitness, whereas genotypes with more than $k$ mutations are lethal and
have fitness zero \cite{Wolff2009}. Hence the fitness landscape is defined as
\begin{equation}
	w_\sigma=
	\begin{cases} 1, & \text{ if } d_\sigma \le k,\\
		0, & \text{ otherwise,}
	\end{cases}
	\label{Eq:mesa_land}
\end{equation}
where $d_\sigma$ is the Hamming distance to the wild-type sequence
$(0,0,...,0)$ or, equivalently, the number of loci with
allele $1$. We will refer to $k$ as the mesa width
or as the critical Hamming distance. 

\begin{figure}%[h!]
	\centering
	\includegraphics[width=0.6\linewidth]{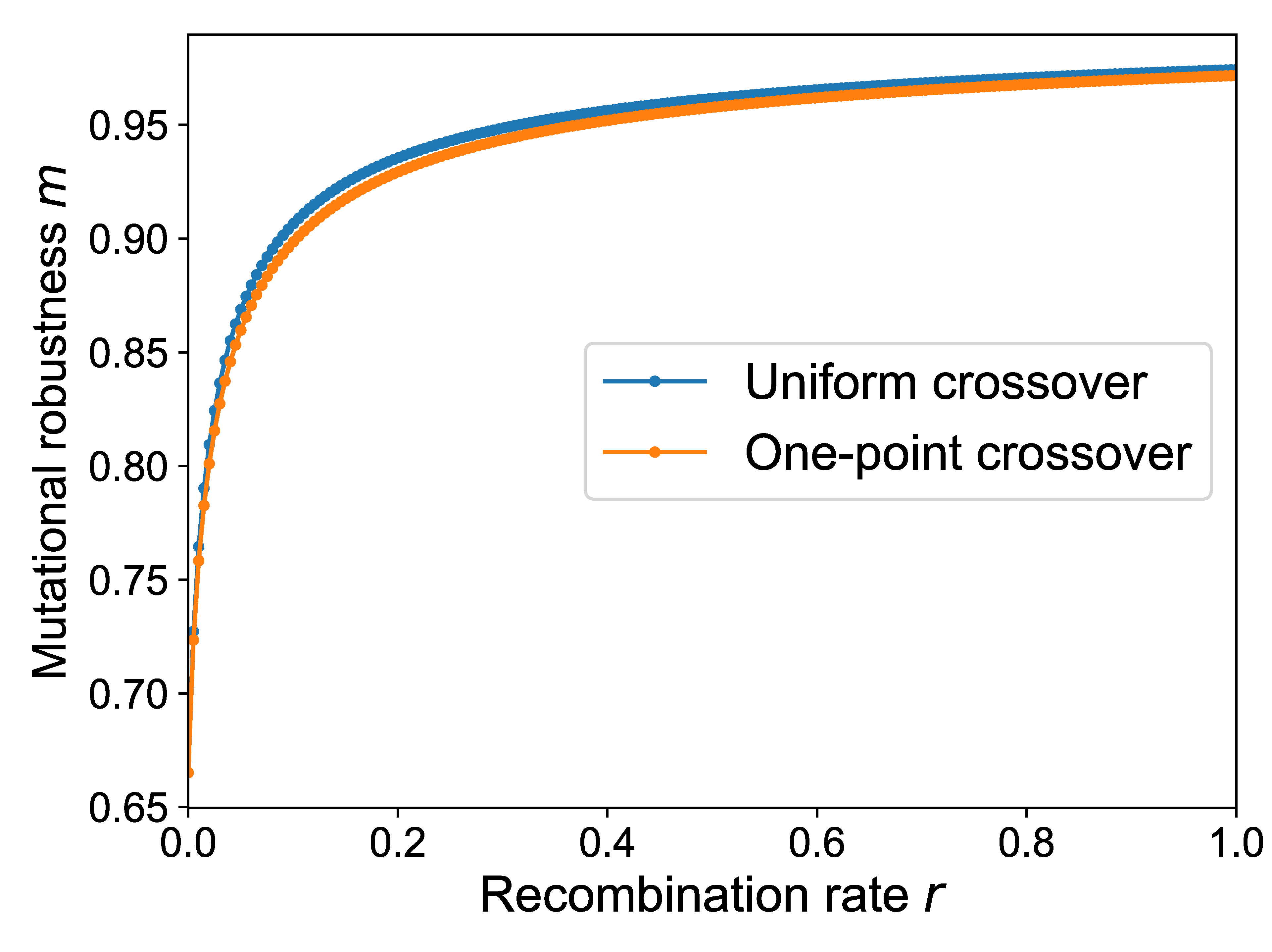}
	\caption[Source1]{\textbf{Mutational robustness in a mesa
			landscape as a function of recombination rate.} Data points
		are obtained by numerically iterating the
		selection-mutation-recombination dynamics until the
		equilibrium state is reached. The parameters of the mesa
		landscape are $L=6$, $k=2$ and the mutation rate is $\mu=0.001$.}
	\label{fig:m_rmesa}
\end{figure}

Such a scenario can for example be observed in
the evolution of regulatory motifs, 
where the fitness depends on the binding affinity of the regulatory
proteins and $d_\sigma$ corresponds to the number of mismatches compared
to the original binding motif \cite{Gerland2002,Berg2004}. 
The two-locus model discussed in the preceding section corresponds to
the mesa landscape with critical Hamming distance $k=1$ and sequence
length $L=2$. 
Here we ask to what extent the behavior observed for the two-locus
model generalizes to longer sequences and variable $k$. Numerical
simulations suggest that the strong increase of mutational robustness
with recombination rate indeed persists in the general setting, and the
particular recombination scheme seems to have only a minor influence; see Fig~\ref{fig:m_rmesa}.  

Whereas an analytical treatment for general $L, k$ and intermediate
recombination rates appears to be out of reach, accurate
approximations are available in the limiting case of strong recombination
or of no recombination, assuming mutation rate is small. The full derivations for both cases can
be found in S1~Appendix. In the following we summarize the main
results. 

\paragraph*{Strong recombination.} In the limit of strong
recombination  we demand linkage equilibrium after each recombination
step. This is satisfied if we use the so-called communal recombination
scheme \cite{Neher2010}. In this scheme an individual is not the
offspring of a pair of parents. Rather, its genotype is aggregated by
choosing the allele at each locus from a randomly selected parent. 
Hence the probability of occurrence
of an allele at each locus in the offspring genotype after recombination is given by
the corresponding allele frequency of the whole population, which is precisely 
the definition of linkage equilibrium. 
In order to obtain an approximation for the mutational robustness we
further assume that the mutation rate $\mu$ is small, which in turn
implies a low frequency of mutant alleles. 
Following the derivation in S1~Appendix this leads us to the expression 
\begin{equation}
	m_\textrm{cr} \approx 1 - {\binom{L-1}{k}}^{1/(k+1)}\mu^{k/(k+1)}+ \frac{k}{k+1}\mu,
	\label{Eq:m_srwm}
\end{equation}
which can be approximated as
\begin{equation}
	m_\textrm{cr} \approx 1 - U^{k/(k+1)} {(k!)}^{-1/(k+1)} + \frac{k}{k+1}\mu
	\label{Eq:m_srwm2}
\end{equation}
for $L \gg k$, where $U = L \mu$ is the genome-wide mutation rate and the subscript signifies the communal recombination scheme.
Using Eq~\eqref{Eq:m_srwm} and setting $L=2$ and $k=1$ we retrieve
the result \eqref{Eq:mr1} for the two-locus model. Furthermore
comparing Eq~\eqref{Eq:m_srwm} and Eq~\eqref{Eq:m_srwm2} to
numerical simulations of communal recombination illustrates their
validity for large $L$ (S2~Fig). 
If we use uniform crossover and one-point crossover instead of
communal recombination, the numerical simulations suggest that the
leading behaviour of $1-m$ is still a function of $U = L\mu$ with the same
exponent $k/(k+1)$, which supports the universality of our findings
with respect to the recombination scheme; see S3~Fig.

\paragraph{No recombination.}
In order to obtain analytical results in the absence of recombination
we assume that the mutation rate is small enough that only a
single point mutation occurs in one generation. This condition is
fulfilled if $U=L\mu \ll 1$. Interestingly, we observe that in this
regime the equilibrium frequencies after selection are independent of
$U$. Therefore also the mutational robustness after selection, denoted
by $M_\textrm{nr}$, is independent of $U$. The relation between mutational robustness after selection ($M_\textrm{nr}$) and after mutation ($m_\textrm{nr}$) is given by 
\begin{equation}
	m_\textrm{nr}=M_\textrm{nr}(1-U)+M_\textrm{nr}^2 U,
\end{equation}
which makes it suffice to find $M_\textrm{nr}$. 

Assuming $k/L \ll 1$ it is
possible to link the set of stationarity conditions to the Hermite
polynomials $H_n(x)$. This yields an approximation for the mutational
robustness after selection as
\begin{equation}
	M_\textrm{nr}=\sqrt{\frac{y_k}{L}}+o(L^{-1/2}),
\end{equation}
where $\sqrt{y_k/2}$ is the largest zero of
$H_{k+1}(x)$. Correspondingly, the mutational robustness after mutation is
\begin{equation}
	m_\textrm{nr}=\sqrt{\frac{y_k}{L}}(1-U)+\frac{y_k}{L}U.
\end{equation}
A comparison to the exact solutions for $M_\textrm{nr}$, which have been
obtained up to $k=4$, confirms this approximation. 
If we further assume that $1 \ll k \ll L$, we find $y_k \sim 4k$, which leads to
\begin{equation}
	m_\textrm{nr}=2\sqrt{\frac{k}{L}}(1-U)+4\frac{k}{L}U.
	\label{Eq:m_nrwm_x0small}
\end{equation}

Results for the joint limit $k, L \to \infty$ at fixed ratio $x=k/L$
can be obtained from the analysis of Ref.~\cite{Wolff2009}, which yields
\begin{align}
	\label{Eq:mesa_nr_M}
	M_\textrm{nr} = 
	\begin{cases}
		2 \sqrt{x(1-x)}, & \text{ if } x < 1/2,\\
		1, & \text{ if } x \ge 1/2
	\end{cases}
\end{align}
and therefore
\begin{align}
	\label{Eq:mesa_nr}
	m_\textrm{nr} = \begin{cases}
		2 \sqrt{x (1-x)} \left(1-U\right) + 4 x (1-x)
		U, & \text{ if } x < 1/2,\\
		1, & \text{ if } x \ge 1/2.
	\end{cases}
\end{align}
The leading behaviour for small $x$ coincides with
Eq~\eqref{Eq:m_nrwm_x0small}. A comparison of the approximations to
numerical solutions is given in S4~Fig.

\paragraph{Comparison of the two cases.}
It is instructive to compare the results obtained above to the
mutational robustness $m_0$ of a uniform population distribution. 
For the latter we assume that all viable genotypes have the same frequency and all lethal genotypes have frequency zero. For the mesa model this yields
\begin{equation}
	m_{0}(L,k)=\frac{1}{\sum_{i=0}^{k}\binom{L}{i}}\left[\binom{L}{k}\frac{k}{L}+\sum_{i=0}^{k-1}\binom{L}{i}\right]
	\approx \min [2k/L, 1], 
	\label{Eq:mesa_m0}
\end{equation}
where the last approximation is valid for $L \to \infty$.
In S5~Fig the behavior of $m_0$, $m_\textrm{nr}$ and
$m_\textrm{cr}$ is depicted as a function of various model
parameters. Similar to the results obtained for the two-locus model,
we see that selection gives rise to a moderate increase of robustness (from $2k/L$
to $2 \sqrt{k/L}$ for $1 \ll k \ll L$), but recombination has a much
stronger effect and leads to values close to the maximal robustness $m
= 1$ for a broad range of conditions. 

\begin{figure}%[h!]
	\centering
	\includegraphics[width=1.0\linewidth]{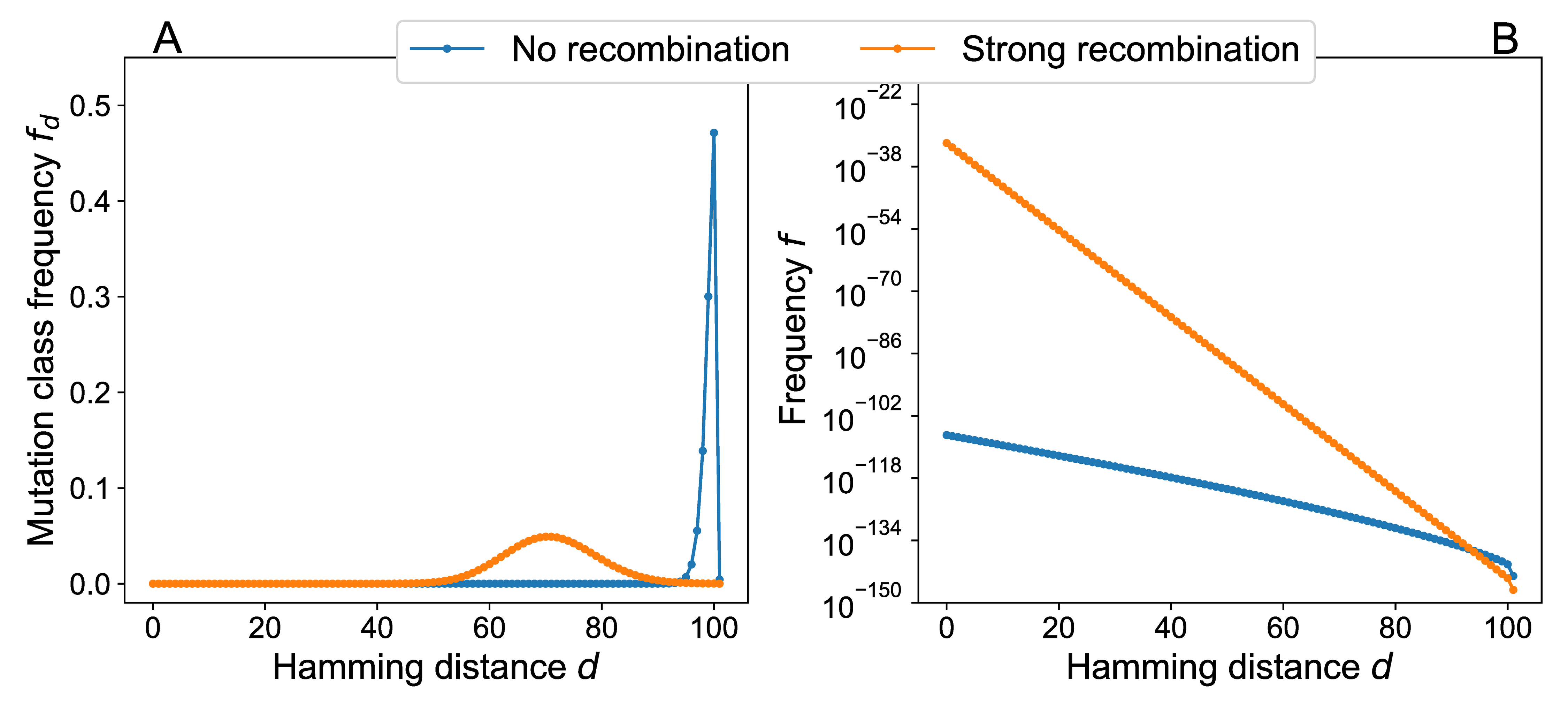}
	\caption[Source1]{
		\textbf{Equilibrium genotype distributions in a mesa landscape for strongly and non-recombining populations.}
		Stationary states for populations with communal
		recombination and no recombination have been computed by assuming
		that only single point mutations occur with
		$U=0.01$. Landscape parameters are $L=1000$ and
		$k=100$.  The resulting mutational robustness is
		$m_\textrm{nr}\approx0.572$ for the non-recombining
		population and $m_\textrm{cr}\approx 1.000$ for
		communal recombination.
		(A) Lumped mutation class frequencies on linear
		scales. In the absence of recombination the majority
		of the population is located at the critical
		Hamming distance $d=k$, whereas in the case of strong
		recombination the distribution is broader and shifted
		away from the brink of the mesa. (B) Genotype
		frequencies on semi-logarithmic scales. In both cases
		the genotype frequencies decrease exponentially with
		the Hamming distance to the wild type, but the
		distribution has much more weight at small distances
		in the case of recombination. 
	}
	\label{fig:m_mesa_sr_nr}
\end{figure}

To elucidate the underlying
mechanism, it is helpful to consider the shape of the equilibrium
frequency distributions in genotype space
(Fig~\ref{fig:m_mesa_sr_nr}). The combinatorial increase of the
number of genotypes with increasing $d_\sigma$ generates a strong
entropic force that selection alone cannot efficiently counteract. As
a consequence, the non-recombining population distribution is
localized near the brink of the mesa at $d_\sigma = k$
\cite{Wolff2009}. In contrast, the contracting property of
recombination \cite{Szollosi2008} allows it to localize the population
in the interior of the fitness plateau where most genotypes are
surrounded by viable mutants. 

S6~Fig shows the
corresponding recombination weight profile. Similar to the genotype
frequencies in Fig~\ref{fig:m_mesa_sr_nr}(B) the recombination weight decays
rapidly with increasing Hamming distance for $r > 0$, but the decay
appears to be faster than exponential. Interestingly, at $d=k$ the
recombination weight decreases with increasing $r$ [see also
Eq~\eqref{eq:2locus_weights}]. The method used to compute
$\lambda_\sigma$ for large mesa landscapes is explained in S1~Appendix.

\begin{figure}%[!h]
	\centering
	\includegraphics[width=0.7\linewidth]{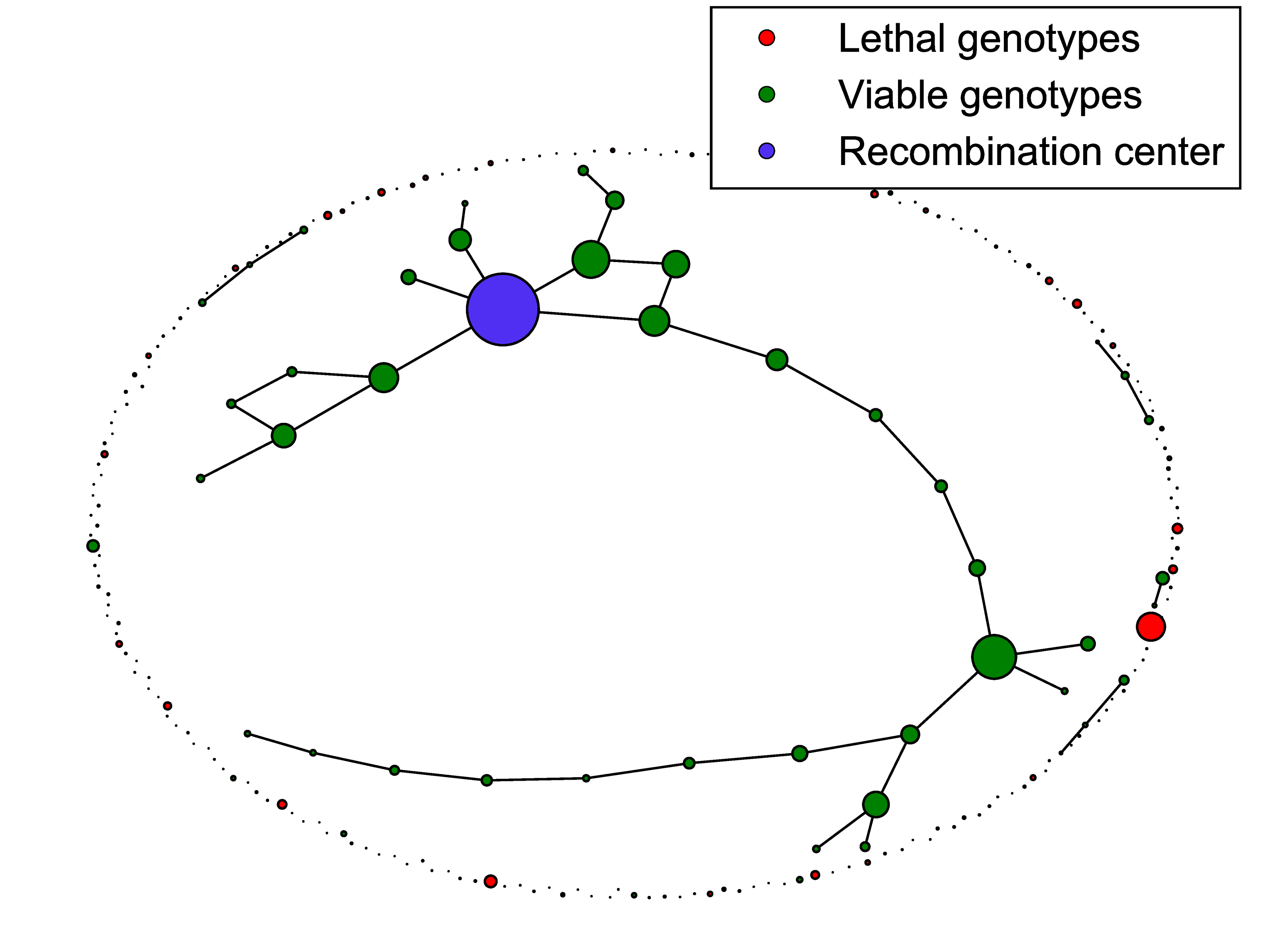}
	\caption[Source1]{\textbf{Network representation of a
			percolation landscape}. The figure shows a percolation
		landscape with $L=8$ loci and a fraction $p=0.2$ of viable
		genotypes. Viable genotypes at Hamming distance $d=1$ are
		connected by edges, and the node area of a genotype
                $\sigma$ is
		proportional to $\lambda_\sigma^6$, where the
                recombination weight $\lambda_\sigma$ is defined in Eq~\eqref{eq:recombweight}. The recombination center is the genotype with the
		largest recombination weight.}
	\label{fig:p1}
\end{figure}

\subsection*{Percolation landscapes}
In the percolation landscape genotypes are randomly chosen to be viable
($w_\sigma = 1$)
with probability $p$ and lethal ($w_\sigma = 0$) with probability $1-p$.
An interesting property of the percolation model is the emergence of
two different landscape regimes
\cite{Gavrilets1997,Gavrilets1997a,Reidys1997,Reidys2009}. 
Above the percolation threshold $p_c$, viable genotypes connected by
single mutational steps form a cluster that extends over the whole
landscape, whereas below $p_c$ only isolated small clusters
appear. Since the percolation threshold depends inversely on the
sequence length, $p_c \approx \frac{1}{L}$, for large $L$ a small
fraction of viable genotypes suffices to create large neutral networks.
This allows a population to evolve to distant genotypes
without going through lethal regions, and correspondingly the
percolation model is often used to study speciation
\cite{Gavrilets2004,Gavrilets1997}. A network representation of the
percolation model is shown in Fig~\ref{fig:p1}. The algorithm used to
generate this visual representation is explained in S1~Appendix.  

\begin{figure}
	\centering
	\includegraphics[width=1.0\linewidth]{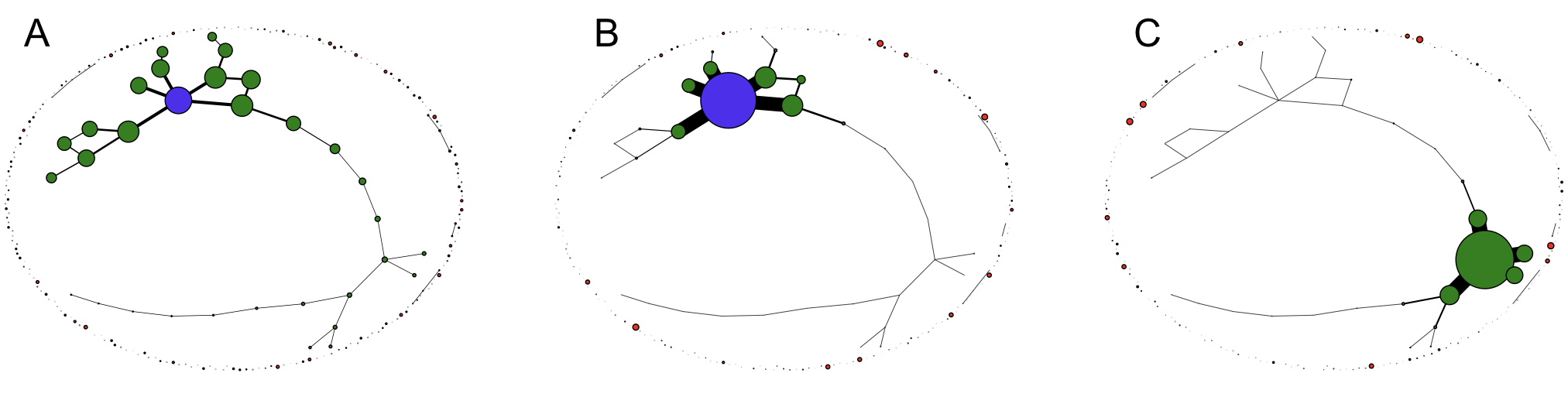}
	\caption[Source1]{\textbf{Stationary states in a percolation
			landscape.} The figure shows three different stationary
		population distributions in the percolation landscape
		depicted in Fig~\ref{fig:p1}.
		Node areas are proportional to the stationary frequency of the respective genotype
		in the population, and the edge width $e_{\sigma,\tau}$ between
		neighboring genotypes is proportional to the frequency
                of the more populated one, $e_{\sigma, \tau} \propto
                \max[f^\ast_{\sigma}, f^\ast_{\tau}]$. (A) Unique stationary state of a
		non-recombining population. (B,C) Stationary states for
		recombining populations undergoing uniform crossover with
		$r=1$. The recombination center (purple) is the most populated genotype
		in (A,B), but not in (C). In all cases the mutation rate is $\mu=0.01$.}
	\label{fig:p2}
\end{figure} 
Fig~\ref{fig:p2} shows three exemplary stationary genotype
frequency distributions on the landscape depicted in
Fig~\ref{fig:p1}.  
In the absence of recombination the equilibrium frequency distribution
is unique, but in the presence of recombination the non-linearity of the
dynamics implies that multiple stationary states may exist
\cite{Paixao2015,Higgs1998,Park2011}. Fig~\ref{fig:p2} displays two
stationary distributions for $r=1$ which are accessed from different
initial conditions. It is visually apparent that the recombining
populations are concentrated on a small number of highly connected
genotypes, leading to a significant increase of mutational robustness.  

To quantify this effect, the average mutational robustness
$\overline{m}$ is calculated as a function of the recombination rate
according to the following numerical protocol:

\begin{itemize}
	\item A percolation landscape for given $L$ and $p$ is generated and
	the initial population is distributed uniformly among all
	genotypes. 
	
	\item The population is evolved in the absence of recombination
	($r=0$) until the unique equilibrium frequency distribution is
	reached, for which the mutational robustness $m$ is calculated. 
	
	\item Next the recombination rate is increased by predefined
	increments. After increasing $r$, the population is again evolved using the stationary state obtained before the increment of $r$ as the initial condition, until it reaches a new
	stationary state for which the mutational robustness is measured. 
	
	\item When the recombination rate has reached $r=1$, a new percolation
	landscape is generated and the process starts all over again. 
	This is done for an adjustable number of runs over which the average
	is taken.
	
\end{itemize}

\begin{figure}
	\centering
	\includegraphics[width=0.6\linewidth]{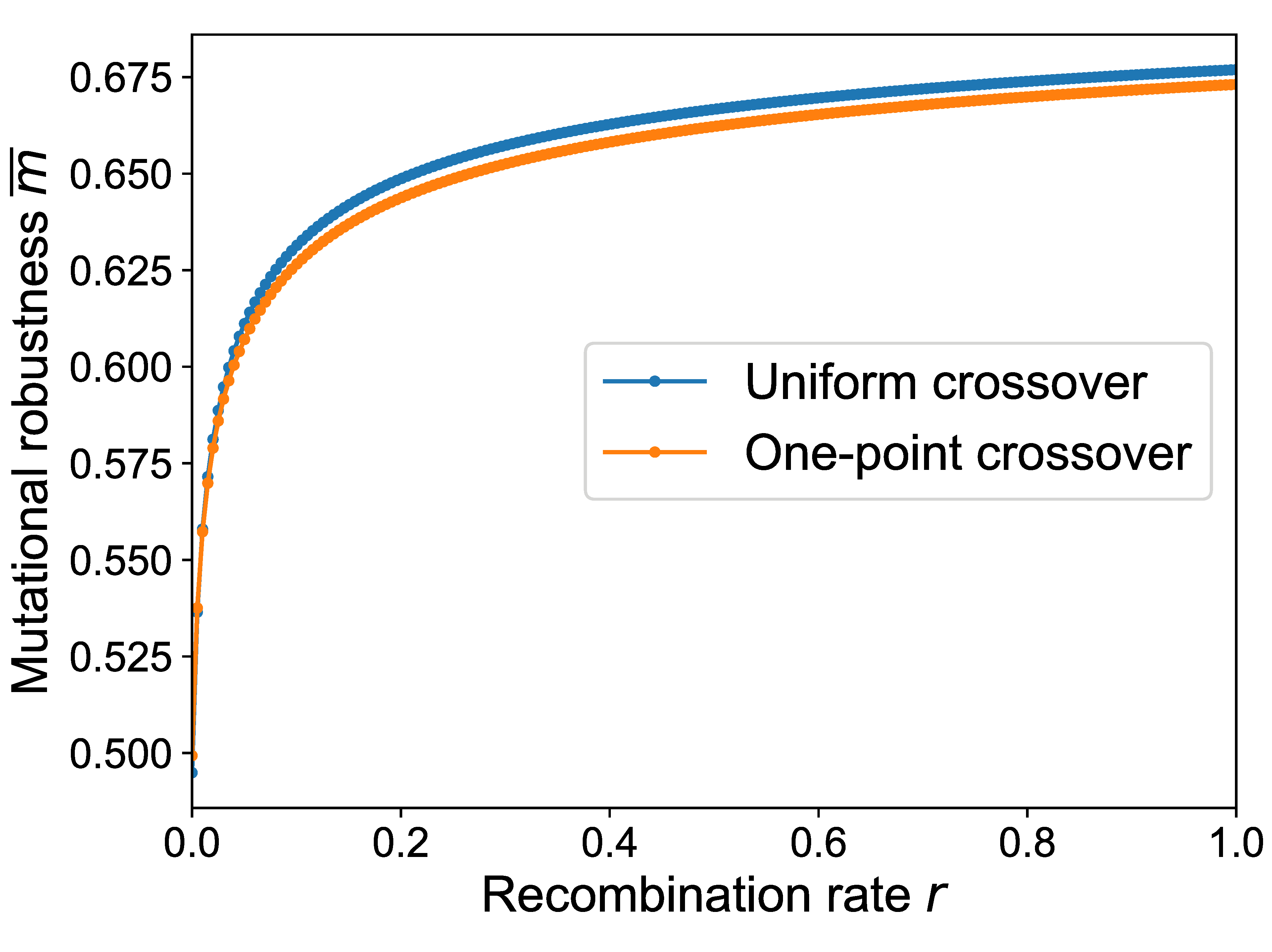}
	\caption[Source1]{\textbf{Average mutational robustness in the percolation landscape as a function of recombination rate.} 
		Mutational robustness is computed for 250 randomly generated
		percolation landscapes with $L=6$ and $p=0.4$, 
		and the results are averaged to obtain $\overline{m}(r)$. The
		mutation rate is $\mu = 0.001$. 
	}
	\label{fig:m_rp}
\end{figure}

The results of such a computation are shown in
Fig~\ref{fig:m_rp}. Similar to the mesa
landscapes, a strong increase of mutational robustness is observed
already for small rates of recombination, and the effect is largely
independent of the recombination scheme. However, in contrast to the mesa landscape the robustness does
not reach its maximal value $m=1$ for $r=1$ and small $\mu$. This
reflects the fact that maximally connected genotypes with $m_\sigma = 1$
are very rare at this particular value of $p$.

For the purpose of comparison we also determined
the average mutational robustness $\overline{m}_0$ of a uniform
population distribution for the percolation model. 
Conditioned on the number $v = \vert V \vert$ of viable genotypes and assuming that $v
\geq 1$, we have
$m_0(v,L) = n(v,L)/L$, where $n(v,L)$ is the average
number of viable neighbors of a viable genotype. The latter is given
by the expression
\begin{equation}
	n(L,v)=\frac{(v-1)L}{2^L-1},
\end{equation}
since for a given viable genotype there are $v-1$ remaining genotypes,
each of which has the probability $L/(2^L-1)$ to be a neighboring
one. Taking into account that the number of viable genotypes is
binomially distributed with parameter $p$ and that the empty hypercube
($v=0$) should yield $m_0=0$ we obtain
\begin{align}
	\label{eq:perc_uni}
	\overline{m}_0 = \sum_{v=1}^{2^L}\frac{(v-1)}{2^L-1}\binom{2^L}{v}p^v{(1-p)}^{2^L-v} = \frac{2^L p-1+{(1-p)}^{2^L}}{2^L-1},
\end{align}
which simplifies to $\overline{m}_0 = p$ when $2^L p \gg 1$.
Note that the condition $2^L p \gg 1$ is naturally satisfied beyond the percolation threshold.

\begin{figure}%[!h]
	\centering
	\includegraphics[width=0.6\linewidth]{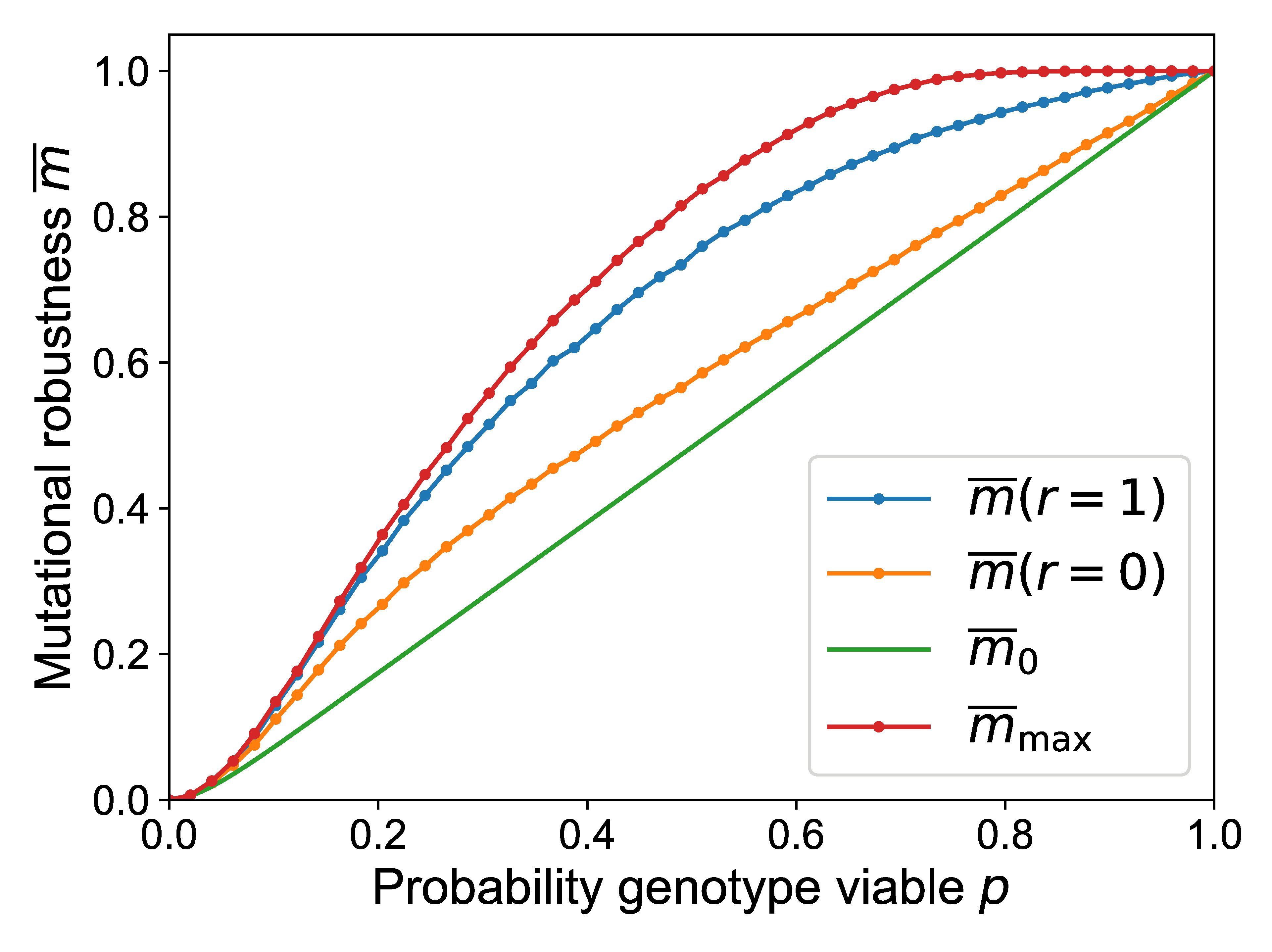}
	\caption[Source1]{\textbf{Mutational robustness in the
			percolation landscape as a function of the fraction of
			viable genotypes.} The robustness for recombining
		($\overline{m}(r=1)$) and non-recombining
		($\overline{m}(r=0)$) populations is obtained by averaging
		over 6800 randomly generated landscapes with $L=6$ and $\mu = 0.001$. In the same
		way the average maximal robustness $\overline{m}_\mathrm{max}$
		is estimated. The full line shows the analytic expression
		\eqref{eq:perc_uni} for the robustness of a uniformly distributed
		population.}
	\label{fig:mutrob_p}
\end{figure}

Fig~\ref{fig:mutrob_p} illustrates that the dynamics induced by
mutation and selection already increase mutational robustness compared
to $\overline{m}_0$ and that the addition of recombination even
further increases mutational robustness for all values of $p$. The
figure also displays the expected maximum number of viable neighbors
of any genotype in the landscape, $\overline{m}_\mathrm{max}$, which
provides an upper bound on the robustness. 
The fact that the numerically
determined robustness remains below this bound for all $p$ shows that
the ability of recombination to locate the most connected genotype is
limited. In S1~Appendix it is shown that
$\lim_{L \to \infty} \overline{m}_\mathrm{max} = 1$ for $p > \frac{1}{2}$.

As outlined above,
the algorithm used to generate Figs~\ref{fig:m_rp} and \ref{fig:mutrob_p} computes the mutational robustness of a particular stationary frequency 
distribution of the recombining population which is smoothly connected to the unique non-recombining stationary state. Although one expects this state to be representative in the sense of being
reachable from many initial conditions, for large enough $r$ there can be multiple stationary states
that will generally display different robustness (see Fig~\ref{fig:p2}). To illustrate this point, S7 Fig shows the results of a simulation of the percolation model 
where all stationary states were identified using localized initial 
conditions, and the mutational robustness was computed separately for each state. 
Whereas on average the mutational robustness is always enhanced by recombination, there are rare instances when recombination reduces the robustness compared to the non-recombining case. 
This may happen, for example, if recombination traps the population on a small island of viable genotypes \cite{deVisser2009,Nowak2014,Park2011,Altland2011}.

\subsection*{Sea-cliff landscapes}
In this section we introduce a novel class of fitness-landscape models 
(to be called sea-cliff landscapes)
that interpolates between the mesa and percolation landscapes. 
Similar to the mesa landscape, the fitness values of the sea-cliff model are 
determined by the distance to
a reference genotype $\kappa^*$. The model differs from the mesa landscape in that it is not assumed that
all genotypes have zero fitness beyond a certain number of
mutations. Instead, the likelihood for a mutation to be lethal (to
``fall off the cliff'') is taken to increase with the Hamming distance from the reference genotype. This is
mathematically realized by a Heaviside step function $\theta(x)$
that contains an uncorrelated random contribution $\eta_\sigma$ and the distance
measure $d(\sigma, \kappa^*)$,
\begin{align}
	w_\sigma=\theta[\eta_\sigma-d(\sigma, \kappa^*)] = \begin{cases}
		1, & \text{ if }\eta_\sigma > d(\sigma,\kappa^\ast),\\
		0, & \text{ if } \eta_\sigma < d(\sigma,\kappa^\ast).
	\end{cases}
\end{align}
This construction is similar in spirit to the definition of the
Rough-Mount-Fuji model \cite{Aita2000,Neidhart2014}. 

The average shape of the landscape can be tuned by the mean $c$ and the standard
deviation $s$ of the distribution of the random variables
$\eta_\sigma$,  which we assume to be Gaussian in the following. 
The average fitness at distance $d$ from the reference sequence is
then given by 
\begin{equation}
	\overline{w}(d) = \text{Prob}(w_\sigma=1) = \frac{1}{2} \left[1-\text{erf}\left(\frac{d-c}{s \sqrt{2}}\right)   \right],
\end{equation}
where $\text{erf}(x)$ is the error function.
Note that the mesa landscape is reproduced if we take the limit $s \rightarrow 0$ 
for fixed $c$ in the range $k < c < k+1$ and the percolation landscape
is reproduced if we take a joint limit
$s,|c| \rightarrow \infty$ with $c/s$ fixed.

To fix $c$ and $s$ we introduce two distances $d_<$ and and $d_>$ such
that  
$\overline{w}(d_<) = 0.99$ and 
$\overline{w}(d_>) = 0.01$,
which leads to the relations
\begin{equation}
	c = \frac{1}{2}(d_< + d_>) \quad \text{and} \quad s \approx 0.215(d_>-d_<). 
\end{equation}
The model can be generalized to include several predefined reference sequences,
\begin{equation}
	w(\sigma)=\theta\left \{ \sum_{\kappa^*} \theta[\eta_{\sigma,\kappa^*}-d(\sigma, \kappa^*)]\right \},
\end{equation}
which allows to create a genotype space with several highly connected
clusters. Depending on the Hamming distance between the reference
sequences and the variables $c$ and $s$, clusters can be isolated or
connected by viable mutations.

\begin{figure}%[!h]
	\centering
	\includegraphics[width=1.0\linewidth]{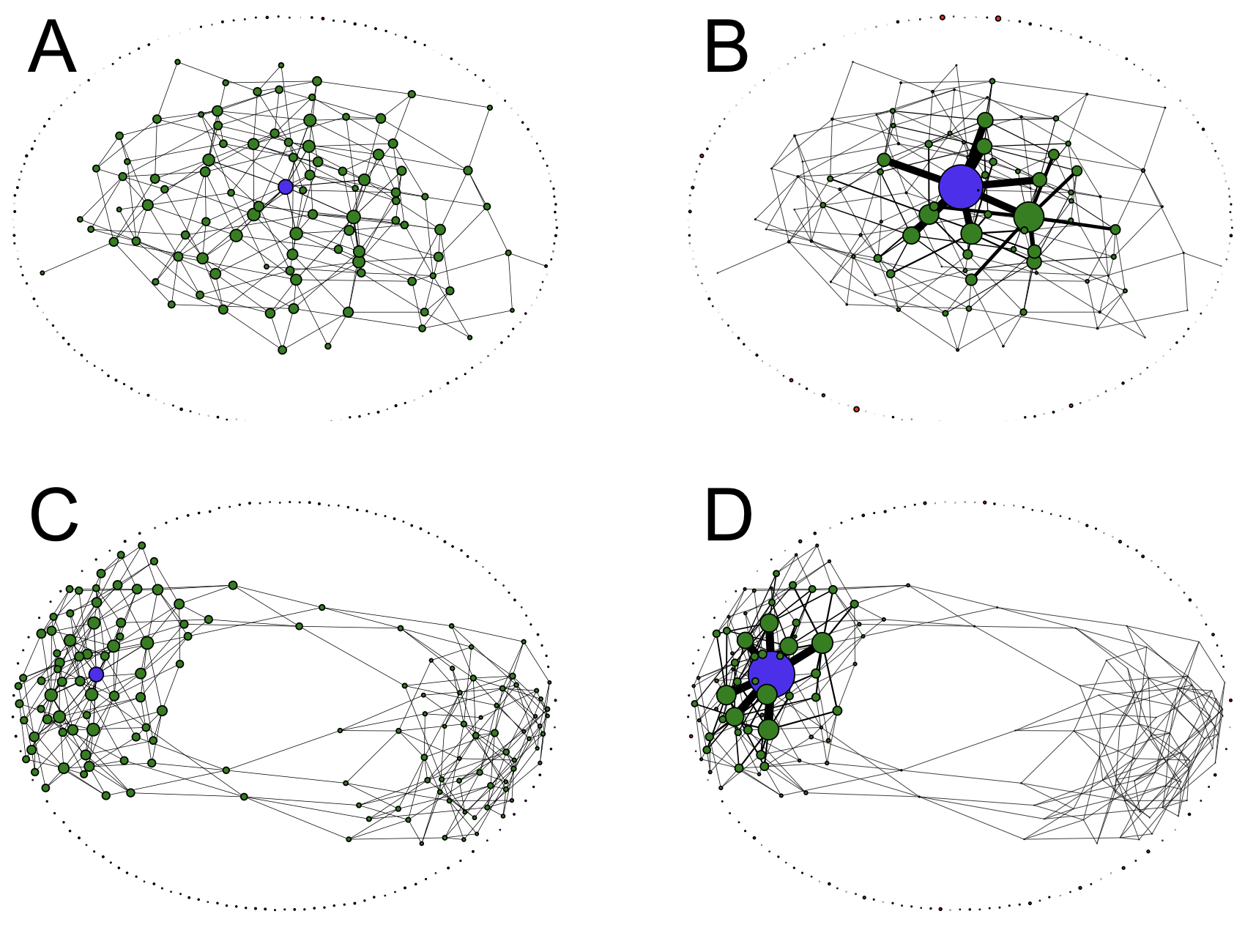}
	\caption[Source1]{
		\textbf{Stationary states in two different sea-cliff landscapes with and without recombination.}
		(A,B) A single reference genotype with landscape
                parameters $L=8$, $d_<=1$ and $d_>=6$. (C,D) Two
                reference genotypes which are antipodal to each other
                with landscape parameters $L=8$, $d_<=2$ and
                $d_>=4.2$. (A,C) Stationary frequency distribution in
                the absence of recombination. (B,D) Stationary
                frequency distribution with uniform crossover and $r =
                1$. In all cases node areas are proportional to
                genotype frequencies, and the recombination center is marked in blue. The edge width between
		neighboring genotypes is proportional to the frequency
                of the more populated one. The mutation rate is $\mu=0.01$.
	}
	\label{fig:2sssc}
\end{figure}

Fig~\ref{fig:2sssc} shows stationary states in the absence and presence of 
recombination for two different sea-cliff landscapes with one and two reference 
genotypes, respectively. Similar to the other landscape models, mutational robustness 
increases strongly with recombination, due to a population concentration within 
a neutral cluster. In the presence of two reference genotypes the recombining 
population should be concentrated within a single cluster. Otherwise  
lethal genotypes would be predominantly created through recombination of genotypes on 
different clusters. This observation can also be interpreted in the context of 
speciation due to genetic incompatibilities\cite{Paixao2015,Gavrilets1997}.
Without recombination genotypes on both clusters have a nonvanishing frequency, 
but still the larger cluster is more populated. In contrast to the percolation 
landscape, robustness reaches a value close to unity for large $r$, 
because highly connected genotypes are abundant close to the 
reference sequence (S8~Fig).

\subsection*{Mutational robustness and recombination weight}

Comparing Figs~\ref{fig:m_rmesa}, \ref{fig:m_rp} and S8~Fig, the dependence of
mutational robustness on the recombination rate is seen to be
strikingly similar. Despite the very different landscape topographies, in all cases a small amount of
recombination gives rise to a massive increase in robustness compared
to the non-recombining baseline. For the mesa landscape this effect
can be plausibly attributed to the focusing property of recombination,
which counteracts the entropic spreading towards the
fitness brink and localizes the population inside the plateau of viable
genotypes. In the case of the holey landscapes, however, it is not
evident that focusing the population towards the center of its
genotypic range will on average increase robustness, since viable and
lethal genotypes are randomly interspersed. 

\begin{figure}%[!h]
	\centering
	\centering
	\includegraphics[width=0.9\linewidth]{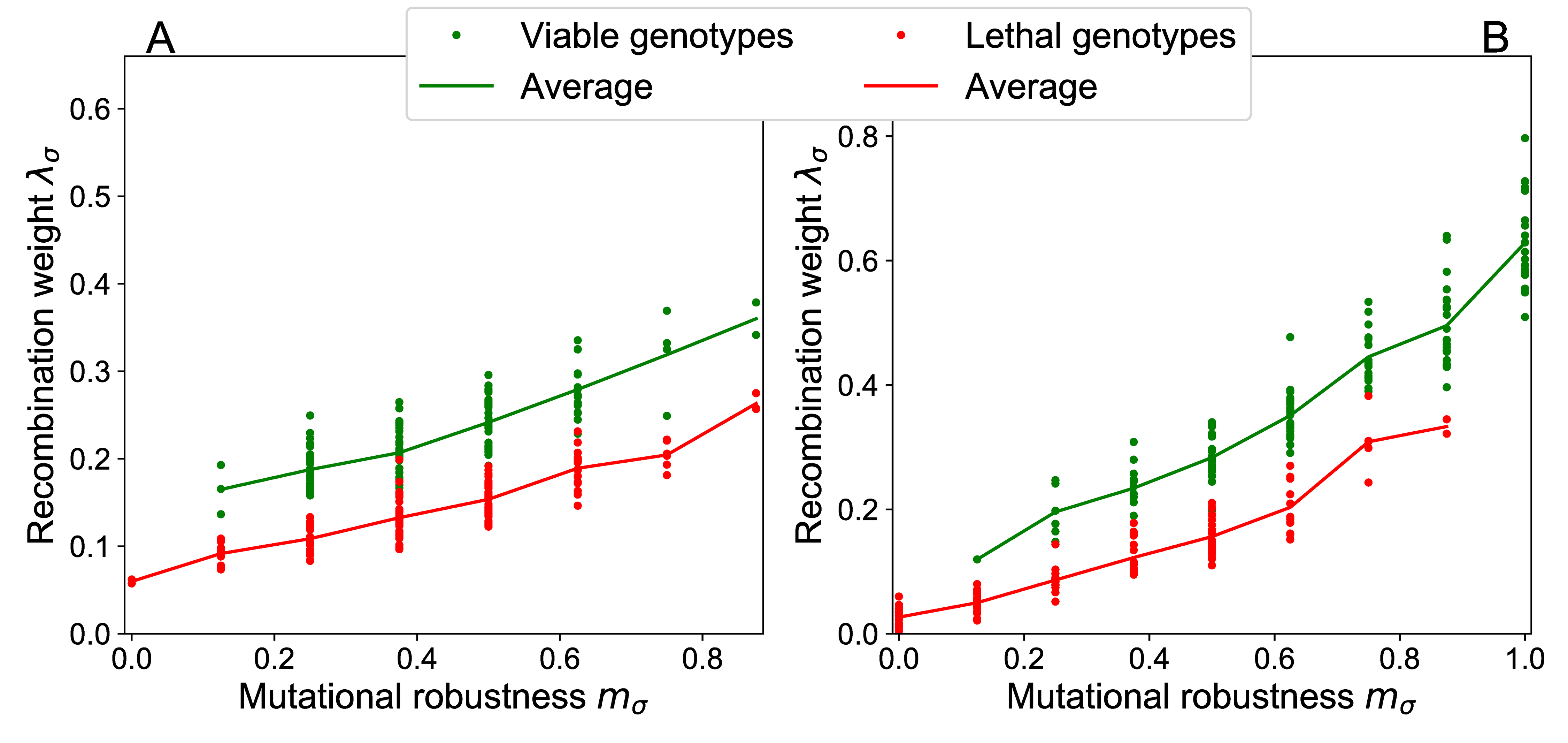}
	\caption[Source1]{\textbf{Mutational robustness correlates
			with recombination weight.} The recombination weight of
		genotypes is plotted against their mutational robustness for
		(A) a percolation landscape with parameters $L=8$, $p=0.4$ and
		(B) a sea-cliff landscape with parameters $L=8$, $d_< = 2$, $d_>
		= 6$. For the evaluation of
		the recombination weight \eqref{eq:recombweight}, uniform crossover at rate
		$r=1$ is assumed.  
	}
	\label{fig:rcomp1}
\end{figure}

To establish the relation between recombination and mutational robustness on the level of individual genotypes, in Fig~\ref{fig:rcomp1} we plot the recombination weight of each genotype
against its robustness $m_{\sigma}$. 
A clear positive correlation between the two
quantities is observed both for percolation and sea-cliff
landscapes. Additionally we differentiate between viable and
lethal genotypes. In the percolation landscape viable genotypes are
uniformly distributed in the genotype space, which implies that lethal
and viable genotypes have on average the same number of viable
point-mutations. Nevertheless
the recombination weight of viable genotypes is larger. The fitness of a genotype influences its own
recombination weight, because the genotype itself is a possible
parental genotype in the recombination event. 

In non-neutral fitness landscapes the redistribution of the population
through recombination competes with selection responding to
fitness differences, and the generalized definition \eqref{eq:recombweight2} of the
recombination weight captures this interplay. To exemplify the
relation between recombination weight and mutational robustness in
this broader context, we use an empirical fitness landscape for
the filamentary fungus \textit{Aspergillus niger} originally obtained
in \cite{deVisser1997}.
In a nutshell, two strains of \textit{A. niger} (N411 and N890) were
fused to a diploid which is unstable and creates two haploids by
random chromosome arrangement. Both strains are isogenic to each
other, except that N890 has $8$ marker mutations on different
chromosomes, which were induced by low UV-radiation. Through this
process $2^8=256$ haploid segregants can theoretically be created of
which $186$ were isolated in the experiment. As a result of a
statistical analysis it was concluded that the missing $70$
haploids have zero fitness \cite{Franke2011}. 

\begin{figure}%[!h]
	\centering
	\includegraphics[width=1\linewidth]{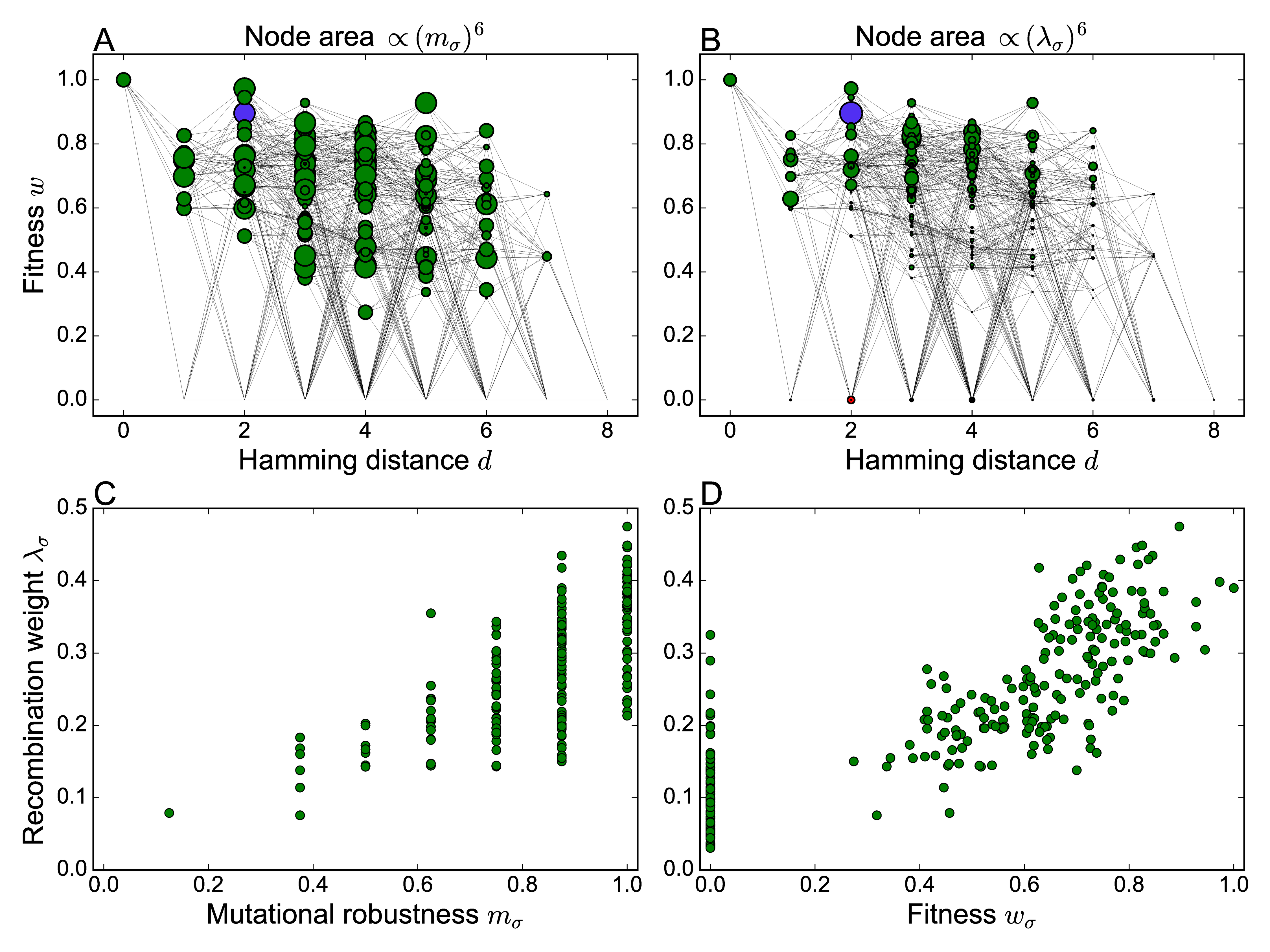}
	\caption[Source1]{\textbf{The empirical \textit{A. niger}
			fitness landscape.} (A,B) Two-dimensional network
		representation of the fitness landscape with node sizes
		determined by the mutational robustness $m_\sigma$ and the
		recombination weight $\lambda_\sigma$, respectively. In
		order to make the differences between genotypes more
		conspicuous, the node area is chosen 
		proportional to the sixth power of these quantities.  
		The recombination weight is evaluated for uniform
		crossover with $r=1$, and the recombination center is
		highlighted in purple. (C,D) Recombination weight plotted
		against mutational robustness and genotype fitness,
		respectively. Lethal genotypes with $w_\sigma = 0$ appear only
		in panel D. 
	}
	\label{fig:AN1}
\end{figure}

In order to illustrate the fitness landscape, a network representation
is employed where genotypes are arranged in a plane according to their
fitness and their Hamming distance to the wild type, which in this
case is the genotype of maximal fitness. In
Figs~\ref{fig:AN1}A,B node sizes are adjusted to the recombination weights
and mutational robustness of genotypes, respectively, in order to
display the distribution of these quantities. In accordance with the analyses for neutral
fitness landscapes, a clear correlation between the recombination weights
and mutational robustness is shown in Fig~\ref{fig:AN1}C. Since fitness
values are not binary we further consider the correlation between the
recombination weights and fitness values (Fig~\ref{fig:AN1}D).
The recombination center is one of the maximally robust genotypes with
$m_\sigma=1$, but it is not the fittest within this group. The wild type has maximal fitness but, by comparison, lower robustness ($m_\sigma=7/8$).

%Note that the recombination center coincide with a maximally robust genotype ($m_\sigma=1$) which is not the wild type that has one lethal neighbor ($m_\sigma=7/8$).

\begin{figure}%[!h]
	\centering
	\includegraphics[width=1\linewidth]{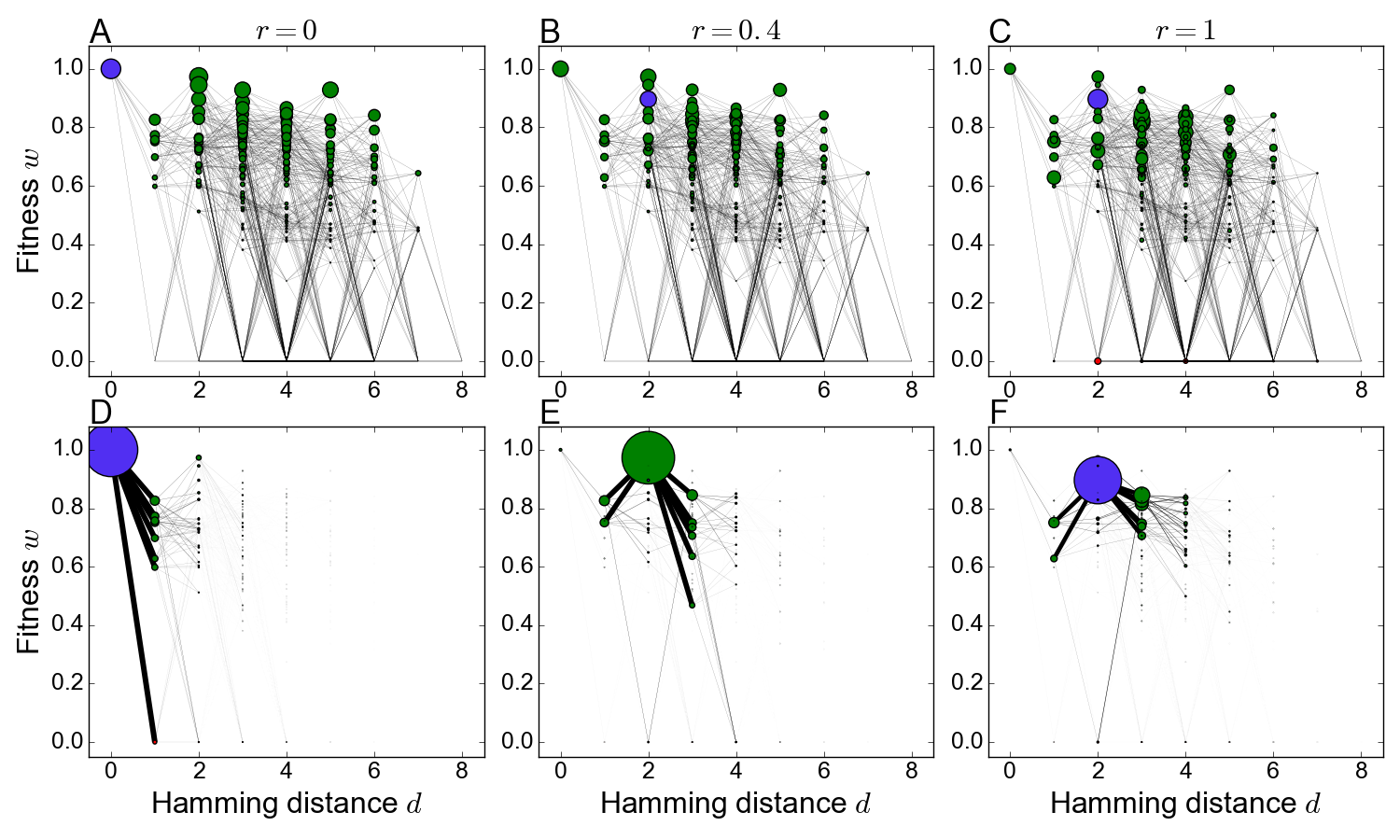}
	\caption[Source1]{\textbf{Recombination weights and stationary
			states at different recombination rates}.(A-C)
		Two-dimensional network representation of the
		\textit{A. niger} fitness landscape with node areas
		proportional to the sixth power of the recombination
		weight for recombination rates $r=0$, $r=0.4$ and $r=1$, respectively. (D-F) Two-dimensional network representation of the
		\textit{A. niger} fitness landscape with node areas
		proportional to the stationary genotype frequency at the
		same recombination rates and mutation rate $\mu = 0.005$. The edge width between
		neighboring genotypes is proportional to the frequency
                of the more populated one.
	}
	\label{fig:AN2}
\end{figure}

Fig~\ref{fig:AN2} highlights how the recombination weights change
as a function of the recombination rate and how this affects the
stationary state of a population. For small recombination rates the
recombination weight of each genotype mainly depends on its own
fitness, and therefore the wild type coincides with the recombination
center. With increasing recombination rate the connectivity of the surrounding genotype network becomes more important and the recombination center
switches to a genotype at Hamming distance $d=2$. In contrast to the
numerical protocol described previously, in the simulations used to
generate Figs~\ref{fig:AN2}D-F the population is reset to a uniform
distribution before the recombination rate is increased. Otherwise the
population would continue to adapt to the wild type, which has the
highest fitness and from which it cannot escape because of peak
trapping \cite{deVisser2009,Nowak2014}.
Starting from an initially uniform distribution the population will
adapt to one of three possible final genotypes which depend on the
recombination rate. For small and large recombination rates the most
abundant genotype coincides with the recombination center (Figs~\ref{fig:AN2}D
and F), whereas for intermediate recombination rates the population
chooses another genotype that is also located at Hamming distance
$d=2$ but has higher fitness (Fig~\ref{fig:AN2}E).
The recombination center ultimately dominates the
population, not only because it is maximally connected ($m_\sigma = 1$), but also because the genotypes that it is
connected to have high fitness. In this sense the sequence of
transitions in the most abundant genotype that occur with increasing
recombination rate is akin to the scenario described previously in
non-recombining populations as the
``survival of the flattest'' \cite{Wolff2009,Wilke2001a}. Along this
sequence mutational robustness increases monotonically whereas the
average fitness of the population actually declines
(S9~Fig). 

\section*{Discussion}
\label{ch:discussion}
Despite a century of research into the evolutionary bases of
recombination, a general mechanism explaining the ubiquity of
genetic exchange throughout the domains of life has not been
found \cite{Otto2002,deVisser2007}. Even within the idealized scenario of a population evolving in a
fixed environment, whether or not recombination speeds up adaptation
and leads to higher fitness levels depends in a complicated way on the
structure of the fitness landscape and the parameters of
the evolutionary dynamics
\cite{Kondrashov2001,deVisser2009,Misevic2009,Moradigaravand2012,Moradigaravand2014,Nowak2014}. 

The most important finding of the present work is that, by comparison,
the effect of recombination on mutational robustness is much simpler
and highly universal. Irrespective of the number of loci, the
structure of the fitness landscape or the recombination scheme,
recombination leads to a significant increase of robustness that is
usually much stronger than the previously identified effect of
selection \cite{vanNimwegen1999,Bornberg1999,Wilke2001}. This suggests that the
evolution of recombination may be closely linked to the evolution of
robustness, and that similar selective benefits are involved in the two
cases. Although the relation of robustness to evolutionary fitness is
subtle and not fully understood \cite{deVisser2003}, it has been
convincingly argued that robustness enhances evolvability and hence
becomes adaptive in changing environments \cite{Wagner2005,Masel2010,Draghi2010,Payne2019}. A common
perspective on recombination, robustness and evolvability can help to develop novel
hypotheses about the evolutionary origins of these phenomena that can
be tested in future computational or empirical studies.

On a quantitative level, we have shown that robustness generally depends
on the ratio of recombination to mutation rates, and that the
robustness-enhancing effect saturates when $r \gg \mu$. This
observation highlights the importance of $r/\mu$ as an evolutionary
parameter. Interestingly, even in bacteria and archaea,
which have traditionally been regarded as essentially non-recombining,
the majority of species displays values of $r/\mu$ that are
significantly larger than one \cite{Guttman1994,Vos2009,Didelot2010}.
Similarly, a recent study of the evolution of \textit{Siphoviridae} phages
revealed a ratio of recombination events to mutational substitutions of about
24 \cite{Kupczok2018}. In eukaryotes this ratio is expected to be
considerably higher \cite{Xia2002}. This indicates that most organisms
maintain a rate of recombination that is sufficient to reap its
evolutionary benefits in terms of increased robustness.

In order to clarify the mechanism through which recombination
enhances robustness, we have introduced the concept of the
recombination weight, which is a measure for the
likelihood of a genotype to arise from the recombination of two viable
parental genotypes. The recombination weight defines a ``recombination
landscape'' over the space of genotypes which is similar in spirit to, but
distinct from, previous mathematical approaches to conceptualizing the
way in which recombining populations navigate a fitness landscape
\cite{Stadler1997}. It is complementary to the more commonly used
notion of a recombination load, which refers to the likelihood for a
viable genotype to recombine to a lethal one \cite{Azevedo2006,Singhal2019}.
In many cases the maximum of the recombination
weight correctly predicts the most populated genotype in a recombining
population at low mutation rate. Moreover, the concept generalizes to
non-neutral landscapes and thus permits to address situations where
selection and recombination compete. 

Provided recombination weight is correlated with mutational robustness for the individual genotypes, this explains the positive effect of recombination on the population-level robustness.
Whether or not such a correlation exists will generally depend on the structure of the fitness landscapes. For simple neutral landscapes such as the mesa landscape it is an immediate
consequence of the focusing property of recombination, but for more complex neutral networks the relationship between the two quantities is nontrivial and needs to be studied on a case-by-case basis.
Although a positive correlation was observed numerically both for the holey landscapes and the empirical landscape considered in this work, it is not difficult to construct landscapes
where the genotypes with high recombination weight are not highly robust. As a simple but instructive example, in S10 Fig we show results for an `atoll' landscape where a ring of viable genotypes surrounds
a central hole of lethals.   

Throughout this work the effects of genetic drift have been neglected. We expect that our results will be applicable to finite populations as long as the population is sufficiently diverse 
rather than being monomorphic. This requires the population-wide mutation rate $N \mu L$ to be much larger than unity \cite{vanNimwegen1999,Szollosi2008}. If $N \mu L \ll 1$ the population is almost always 
monomorphic and recombination has no effect. In this regime the population explores the fitness landscape as a random walker and the observed mutational robustness is the uniform robustness $m_0$. 
In S11 Fig we present the results of finite population simulations on a mesa landscape, which show a sharp transition from the random walk regime to the behavior predicted by the deterministic
theory when $N \mu L \sim 1$.   

Future work should be directed towards extending the present investigation to more realistic genotype-phenotype maps
arising, for example, from the secondary structures of biopolymers
such as RNA or proteins \cite{Huynen1994,Xia2002,Szollosi2008}, or from simple genetic, metabolic or logical
networks \cite{Wagner2005,Azevedo2006,Hu2014,Garcia-Martin2018}. There is ample evidence from numerical studies that a favorable effect of recombination on mutational robustness 
is present also in these more complex systems, but a detailed analysis of the underlying mechanism has not been carried out. This would entail, in particular, the generalization to genotype spaces composed of sequences carrying more than two alleles per site. 
We expect that at least part of the analysis for the mesa landscapes carries over to this setting, and in fact some results for the non-recombining case have already been obtained \cite{Wolff2009}.  
More importantly, the role of the topology of the corresponding neutral networks in shaping the correlation between recombination weight and robustness needs to be explored systematically.
Research along these lines will help to corroborate the relationship
between recombination and robustness that we have sketched, and to further
elucidate the origins of these two pervasive features of biological
evolution.

\section*{Acknowledgments}
JK acknowledges the kind hospitality of the Higgs Centre for Theoretical Physics at the University of Edinburgh during the completion of the manuscript. 
\nolinenumbers

% Either type in your references using
% \begin{thebibliography}{}
% Text
% \end{thebibliography}
%
% or
%
% Compile your BiBTeX database using our plos2015.bst
% style file and paste the contents of your .bbl file
% here. See http://journals.plos.org/plosone/s/latex for 
\bibliography{Klug8}

\section*{Supporting information}

% Include only the SI item label in the paragraph heading. Use the \nameref{label} command to cite SI items in the text.
\paragraph*{S1 Appendix.}
This appendix contains detailed derivations of analytic results
presented in the main text. 

\paragraph*{S1 Fig.}
{\textbf{Population heterogeneity decreases
            with increasing recombination rate.} The figure shows the
          entropy of the genotype frequency distribution in the
          two-locus model defined as $S=-\sum_\sigma f_\sigma^* \ln(f_\sigma^*)$. For
          small mutation rates the strongly recombining population
          primarily consists of a single genotype, which implies that
          $S \to 0$.}

\paragraph*{S2 Fig.}
{\textbf{Mutational robustness for the mesa landscape with communal
            recombination.} The figure compares the analytic
          approximations in Eqs~\eqref{Eq:m_srwm} and \eqref{Eq:m_srwm2} to the
          numerical solution of the stationary genotype frequency distribution
          for the communal recombination scheme. The two panels show
          the mutational robustness as a function of the genome-wide
          mutation rate in linear (A) and double-logarithmic (B)
          scales, respectively.    
		The parameters of the mesa landscape are $L=30$ and
                $k=3$.}
            
\paragraph*{S3 Fig.} 
{\textbf{Mutational robustness in a mesa
		landscape with different recombination schemes.} The figure
	compares the analytic results for communal recombination
	($m_\textrm{cr}$) with numerical data obtained using uniform
	crossover ($m_\textrm{uc}$) and one-point crossover
	($m_\textrm{opc}$) at $r = 1$.
	The landscape parameters are
	$L=5$, $k=2$ and robustness is plotted as a function of the
	genome-wide mutation rate $L \mu$. (A) Mutational robustness
	on linear scales. (B) Double-logarithmic plot of $1-m$
	vs. $L\mu$, illustrating the power-law behavior $1-m \sim
	{(L\mu)}^b$ with the exponent $b = k/(k+1) = 2/3$ predicted
	by the analysis of the communal recombination model.}

 \paragraph*{S4 Fig.}
{\textbf{Mutational robustness for the mesa
            landscape in the absence of recombination.} The figure
          compares the analytic predictions in Eqs~\eqref{Eq:mesa_nr_M}
          and \eqref{Eq:mesa_nr} to the numerical solution for the
          genotype frequency distribution in the absence of
          recombination. The two panels show the mutational robustness
          (A) after selection and (B) after mutation 
          as a function of the scaled mesa width $x_0 = k/L$ for $L=1000$ and $U=0.01$.}

\paragraph*{S5 Fig.}
{\textbf{Mutational robustness in mesa
		landscapes with and without recombination}. Numerical results for communal recombination ($m_\textrm{cr}$) and no recombination ($m_\textrm{nr}$) are shown as dots. The mutational robustness $m_0$ of a uniformly distributed population, given by Eq~(37), as well as the analytic expressions Eqs~(30) and (36) are depicted as lines. (A) Robustness
	as a function of mutation rate $U = L \mu$ for a landscape
	with $L=1000$ and $k= 10$. (B) Robustness as a function of
	mesa width $k$ at fixed $L=1000$ and $U = L\mu = 0.01$. (C)
	Robustness as a function of genome length $L$ at fixed
	$k=10$ and $U = 0.01$. (D) Robustness as a function of genome length $L$ at fixed
	$k=10$ and $\mu = 0.001$.}
 
\paragraph*{S6 Fig.}
{\textbf{Recombination weight in a mesa landscape.} 
	The parameters of the mesa landscape are $L=100$ and
	$k=10$. For $r=0$ the recombination weight is directly
	proportional to the fitness and hence equal for all
	viable genotypes. Already small rates of recombination
	are sufficient to redistribute the recombination weight
	such that the weight of genotypes with small Hamming
	distance is strongly enhanced. Beyond
	$d=20$ the recombination weight is identically zero,
	since the recombinant of two viable genotypes cannot
	carry more than $2 k$ mutations.}
 
\paragraph*{S7 Fig.}
{\textbf{Mutational robustness for different stationary states within a percolation landscape}. 
The figure compares the mutational robustness of non-recombining ($r=0$) and recombining ($r=1$) populations on individual realizations of the percolation model with $L=6$ and three values
of $p$. In order to obtain different stationary states we used localized initial population distributions of the form $f_\tau(0) = \delta_{\tau \sigma}$ for all genotypes with mutational 
robustness $m_{\sigma}\neq0$ and propagated them until stationarity. Since the stationary populations are usually highly concentrated for large $r$ and small $\mu$, this is a natural choice in order 
to access all stationary states. Each data point represents the robustness of the recombining population $m(r=1)$ for a particular stationary state. 
Data points within the same landscape are plotted above the corresponding unique robustness of the non-recombining population $m(r=0)$ and connected by a vertical line. 
The orange crosses show the average over all initial conditions.}

\paragraph*{S8 Fig.} 
{\textbf{Average mutational robustness in the sea-cliff landscape as a function of recombination rate.}
	Mutational robustness is computed for 200 randomly generated
	sea-cliff landscapes with parameters $L=6$, $d_< = 1$ and $d_>
	= 5$, and the results are averaged to obtain
	$\overline{m}(r)$. The mutation rate is $\mu = 0.001$.}

\paragraph*{S9 Fig.}
{\textbf{Mutational robustness and average
            fitness in the empirical \textit{A. niger} fitness landscape.} 
	The mutational robustness and the population-averaged fitness
        in the stationary state are computed as a function of
        recombination rate by evolving the population
        from a uniform initial genotype distribution at mutation rate
        $\mu = 0.005$. Jumps mark changes in the most populated genotype.
	}

\paragraph*{S10 Fig.}
{\textbf{Recombination on an atoll landscape}.
	This landscape is similar to the mesa landscape but includes an inner critical radius within which genotypes are lethal. In this example the inner radius is chosen to be $1$ such that only the wild type is lethal. The outer radius is $2$ and the sequence length is $L=7$. The recombination rate is $r=1$ and the mutation rate is $\mu=0.001$. The frequencies $f_{n}$ of the stationary state at the same Hamming distance $n$ are lumped together. The population is concentrated at distance $1$ which is most robust since only one point mutation is lethal, but the recombination center coincides with the lethal wild type. This example shows that the correlation between recombination weight and mutational robustness depends on the topology of the neutral network.}
 
\paragraph*{S11 Fig.}
{\textbf{Finite population size effects}. The figure shows the mutational robustness in a mesa landscape with parameter $L=6, k=2$ as a function of mutation rate. The finite population 
results were obtained using Wright-Fisher dynamics for $N=1000$ individuals. For small mutation rates such that $N \mu L \ll 1$ the  
	 monomorphic population performs a random walk among viable genotypes, which leads to the uniform mutational robustness $m_0$ given by Eq (37) (green dashed line). 
In this regime recombination cannot have any effect. For $N \mu L > 1$ the robustness rises sharply to the value predicted by the infinite population approach. 
At the maximal mutation rate $\mu=0.5$ the population is uniformly distributed among all (lethal or viable) genotypes after the mutation step and recombination has again no effect.}

\end{document}

\title{Recombination and mutational robustness in neutral fitness
  landscapes: Supplementary appendix}

\author{Alexander Klug}
\affiliation{\rm Institute for Biological Physics, University of Cologne, Cologne, Germany}
\author{Su-Chan Park}
\affiliation{\rm Department of Physics, The Catholic University of Korea,
  Bucheon, Republic of Korea}
\author{Joachim Krug}
\affiliation{\rm Institute for Biological Physics, University of Cologne, Cologne, Germany}

\maketitle

\section{Visualization of fitness landscapes as networks}

In order to visualize random neutral fitness landscapes with more than
two loci we make use of a network representation, 
where genotypes that differ by a single mutation are connected by an
edge. Nodes of the network then represent genotypes, which are
arranged according to a spring layout that is based on a
Fruchterman-Reingold force-directed algorithm \cite{cc17}. To describe
this algorithm briefly, nodes are made to repel each other, which is
counteracted by edges that function as springs. This leads to a
process of spring-force relaxation that arrives at an equilibrium
state which in turn is used for the node positions. The equilibrium
state is characterized by clustering of highly connected regions of
nodes. Therefore this algorithm is only useful if not all nodes have
the same number of edges. Hence edges attached to lethal genotypes
are deleted. This leads to a network in which only viable genotypes
that differ by a single mutation are connected. Lethal genotypes are
off the grid and create a ring of repelled nodes.

\section{Two-locus model with unidirectional mutation}

\begin{figure}
	\centering
	\includegraphics[width=0.5\linewidth]{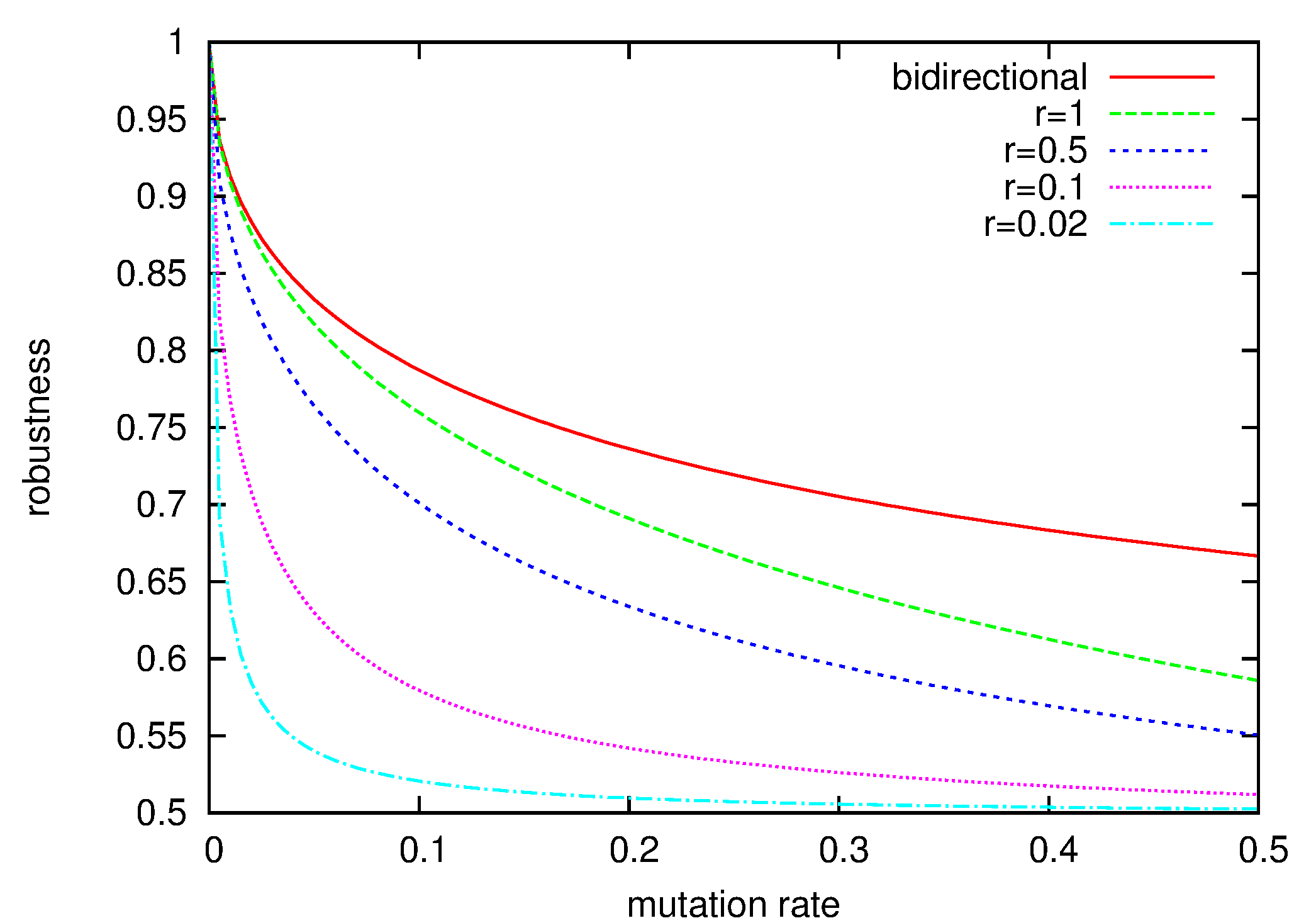}
	\caption[Source1]{\textbf{Mutational robustness in the
            two-locus model with unidirectional mutations}. The figure
          shows the mutational robustness after selection obtained
          for the unidirectional mutation scheme, Eq~\eqref{A1:Muni}, as
          function of $\mu$ for different $r$. For comparison the
          corresponding result Eq~\eqref{A1:Mbi} for the bidirectional
          mutation scheme with $r=1$ is also depicted.} 
	\label{fig:unidir}
      \end{figure}

Following Nowak \textit{et al.} \cite{Nowak1997}, we consider the 
two-locus model with unidirectional mutations from allele 0 to allele 1 at
rate $\mu$ and one-point crossover at rate $r$. Based on the relation 
\begin{equation}
\label{A1:quni}
q_0 = \frac{r}{4 \tilde \mu} q_1^2, \;\;\;\; \tilde \mu = \frac{\mu}{1-\mu}
\end{equation}
between the lumped genotype frequencies after selection, the expression 
\begin{equation}
\label{A1:Muni}
M = q_0 + \frac{1}{2} q_1 = 1 - \frac{\tilde \mu}{r} \left( \sqrt{1 + \frac{r}{\tilde \mu}}
-1 \right)
\end{equation}
can be derived for the mutational robustness after selection. For $r \to 0$ 
this reduces to $M = \frac{1}{2}$ independent of $\mu$, which is smaller
than the value $M = \frac{2}{3}$ expected for a random distribution over
the viable genotypes ($q_0 = \frac{1}{3}, q_1 = \frac{2}{3}$). In the absence 
of recombination, the unidirectional mutations drive the entire population 
into the least robust genotypes (0,1) and (1,0), such that $q_0 = 0$ and $q_1 = 1$.
On the other hand, for $r = 1$ Eq~\eqref{A1:Muni} becomes 
$M = (1+\sqrt{\mu})^{-1}$, which can be compared to the corresponding expression
\begin{equation}
\label{A1:Mbi}
M = \frac{m}{1-f_2} = \frac{2}{2 - \mu + \sqrt{\mu^2 + 4 \mu}}
\end{equation}
obtained from 
Eq~(17) of the main text.
The two expressions coincide
for $\mu \to 0$, but for larger $\mu$ the bidirectional model has higher
robustness, because both selection and recombination contribute to focusing
the population onto the robust genotype (0,0) (Fig~\ref{fig:unidir}).

\section{Mutational robustness on the mesa landscape with communal recombination}
In this section, we calculate the mutational robustness in equilibrium for the mesa landscape,
using the communal recombination scheme~\cite{Neher2010}. Since
fitness depends only on the Hamming distance 
from the wild type, the equilibrium allele-frequency distribution at each locus is the same after
mutation. In the following we denote the (equilibrium) frequency 
of allele $0$ ($1$) after the mutation step  by $\pi_0$ ($\pi_1 = 1 - \pi_0$).
Then the equilibrium frequency $f^\ast_\sigma$ of a genotype $\sigma$ after recombination becomes 
\begin{align}
f^\ast_\sigma = \pi_0^{L-n} \pi_1^n,
\end{align}
where $n$ is the Hamming distance from the wild type. The lumped
frequency of all genotypes in the class with $n$ mutations is then given by
\begin{align}
	f_n = \binom{L}{n} \pi_0^{L-n} \pi_1^n.
\end{align}
Denoting the corresponding lumped frequency after selection by $q_n$
and using the mesa landscape 
defined in Eq~(28) of the main text,
we get 
\begin{align}
	q_n =\begin{cases}f_n \bar w^{-1}, & n \le k, \\ 0,& n > k, \end{cases}
\end{align}
where $\bar w = \sum_{n=0}^k f_n$ is the mean fitness. The lumped
frequency $p_n$ after mutation then satisfies 
\begin{align}
p_d = \sum_{n=0}^L \mu(d|n) q_n,
\end{align}
where $\mu(d|n)$$g_d$ is the probability that a mutation changes the
Hamming distance from $n$ to $d$. The $p_d$ in turn determine the
allele frequency after mutation through
\begin{align}
\pi_1 =\frac{1}{L}\sum_{d=0}^L d p_d
=\frac{1}{L} \sum_d d \sum_n \mu(d|n) q_n
= \frac{1}{L} \sum_n h(n) q_n,
	\label{Eq:p1}
\end{align}
where $h(n) = \sum_d \mu(d|n) d $ is the average Hamming distance of a mutant
generated from a genotype with Hamming distance $n$.
One can easily calculate $h(n)$ for the mutation scheme 
Eq~(4) of the main text,
which yields
\begin{align}
h(n) = n (1-\mu) + (L-n) \mu = L\mu + (1 - 2\mu) n.
	\label{Eq:hn}
\end{align}
This expression has a simple interpretation: On average a fraction
$1-\mu$ of the $n$ mutated sites is not mutated, and a fraction $\mu$
of the $L-n$ non-mutated sites aquires a new mutation.  
Inserting Eq~\eqref{Eq:hn} into Eq~\eqref{Eq:p1}, we finally obtain
\begin{align*}
\pi_1 &= \frac{1}{L} \sum_{n=0}^k \left [ L\mu + (1 - 2\mu) n  \right ] q_n 
= \mu + \frac{1}{L} \left ( 1 - 2 \mu \right ) \sum_{n=0}^k n q_n 
= \mu + \frac{1}{L \bar w} \left ( 1 - 2 \mu \right ) 
\left ( L \pi_1 - \sum_{n=k+1}^L n f_n \right ) \\
&=\mu + \frac{\pi_1}{\bar w} \left ( 1 - 2 \mu \right )
- \frac{1}{L\bar w} \left ( 1 - 2 \mu \right )\sum_{n=k+1}^L n f_n, 
\addtag
	\label{Eq:pfEq}
\end{align*}
where we have used that
\begin{equation}
      \label{eq:sumrule}
      \sum_{n=0}^L n f_n = \sum_{n=0}^L n p_n = L \pi_1,
    \end{equation}
    because the allele frequency is not changed by recombination.

Up to now, everything is exact. It is a formidable, if not impossible, task to find an exact solution
of Eq~\eqref{Eq:pfEq}, so we will solve the problem approximately for small $\mu$.
Since $\mu$ is small, it is plausible to assume that $\pi_1 \ll 1$ as well.
Under this assumption, we can find an approximate expression for $\bar w$ as follows:
\begin{align*}
\bar w &= \sum_{n=0}^k f_n = 1 - \sum_{n=k+1}^L f_n
\approx 1 - \binom{L}{k+1} \pi_1^{k+1} (1 - \pi_1)^{L-k-1} - \binom{L}{k+2} \pi_1^{k+2} ( 1- \pi_1)^{L-k-2}\\
&\approx 1 - \binom{L}{k+1} \pi_1^{k+1} + (L-k-1) \binom{L}{k+1} \pi_1^{k+2} - \binom{L}{k+2} \pi_1^{k+2}\\
&= 1 - \binom{L}{k+1} \pi_1^{k+1} + (k+1) \binom{L}{k+2} \pi_1^{k+2}
\equiv 1 - C_1 \pi_1^{k+1} + (k+1)C_2 \pi_1^{k+2},
\addtag
\end{align*}
where we have kept terms up to order $\pi_1^{k+2}$, $C_i = \binom{L}{k+i}$ ($i=1,2$),
and $1/j!$ should be interpreted as 0 if $j$ is a negative integer. 
Note that the above formula is actually exact for $k \ge L-2$.

Now we approximate Eq~\eqref{Eq:pfEq} term by term. First, we get
\begin{align}
\frac{\pi_1}{\bar w}(1-2\mu)
\approx \pi_1 \left [ 1 + C_1 \pi_1^{k+1} - (k+1)C_2 \pi_1^{k+2}\right ](1-2\mu)
\approx \pi_1 -2 \mu \pi_1 + C_1 \pi_1^{k+2},% - 2 \mu C_1 f_1^{d+2},
\end{align}
where we have kept terms up to $\pi_1^{k+2}$ and $\mu \pi_1$.
Second, we get
\begin{align*}
	\frac{1-2\mu}{\bar w} \sum_{n=k+1}^L n f_n\approx &
	\left [ 1 + C_1 \pi_1^{k+1} - (k+1)C_2 \pi_1^{k+2}\right ](1-2 \mu)\left [ (k+1) C_1 \pi_1^{k+1}(1 - (L-k-1) \pi_1) + (k+2) C_2 \pi_1^{k+2}
\right ]\\
\approx &
(k+1) C_1 \pi_1^{k+1} - k\frac{ L!}{(k+1)!(L-k-2)!} \pi_1^{k+2}.
\addtag
\end{align*}
Accordingly, we arrive at
\begin{align*}
	\pi_1& \approx \mu + \pi_1 - 2 \mu \pi_1 + C_1 \pi_1^{k+2} - \frac{k+1}{L} C_1 \pi_1^{k+1} + k\frac{(L-1)!}{(k+1)!(L-k-2)!} \pi_1^{k+2}\\
&= \pi_1 + \mu - 2 \mu \pi_1 - \binom{L-1}{k} \pi_1^{k+1} + (L-k)\binom{L-1}{k} \pi_1^{k+2},
\addtag
\end{align*}
that is,
\begin{align}
\mu \approx B^{-(k+1)} \pi_1^{k+1} + 2 \mu \pi_1 - (L-k)B^{-(k+1)} \pi_1^{k+2},
\label{Eq:app}
\end{align}
where $B = [k!(L-k-1)!/(L-1)!]^{1/(k+1)}$.
Since the leading behavior of $\pi_1$ is $B \mu^{1/(k+1)}$,
we set
\begin{align}
\pi_1 = B \mu^{1/(k+1)} (1 + g),
\label{Eq:f1}
\end{align}
where $g = o(1)$. Inserting Eq~\eqref{Eq:f1} into Eq~\eqref{Eq:app} and
expanding up to the leading order in $g$, we obtain
\begin{align*}
\mu& \approx \mu (1+g)^{k+1} + 2 B \mu^{(k+2)/(k+1)} - (L-k) B\mu^{(k+2)/(k+1)}\\
&\approx \mu + \mu (k+1) g + (2+k-L) B \mu \mu^{1/(k+1)},
\addtag
\end{align*}
which yields
\begin{align}
g \approx \frac{L-k-2}{k+1}B \mu^{1/(k+1)}.
\end{align}
Therefore the mutational robustness becomes
\begin{align*}
	m&= \sum_{n=0}^{k-1} f_{n} + \frac{k}{L} f_k 
	= 1 - \sum_{n=k+2}^L f_n - \frac{L-k}{L}f_{k}-f_{k+1} 
\approx 1 - \frac{L-k}{L} \binom{L}{k} \left [\pi_1^k- (L-k) \pi_1^{k+1}
\right ] -\binom{L}{k+1} \pi_1^{k+1}\\
&= 1 - \frac{(L-1)!}{k!(L-k-1)!} \pi_1^k + \frac{(L-1)!}{(k+1)!(L-k-1)!}(kL-k^2-k) \pi_1^{k+1}\\
&= 1 - B^{-(k+1)} \pi_1^k + B^{-(k+1)} \pi_1^{k+1} \frac{kL-k^2-k}{k+1}\approx 1 + \mu \frac{kL-k^2-k}{k+1} - B^{-(k+1)}\pi_1^{k+1} \pi_1^{-1}\\
	&\approx 1 + \mu\frac{kL-k^2-k}{k+1} - \mu [1 + (k+1) g] (1-g) \mu^{-1/(k+1)}B^{-1}\approx 1 + \mu\frac{kL-k^2-k}{k+1} - \mu^{k/(1+k)} (1 + k g) B^{-1}\\
	&= 1 + \mu\frac{kL-k^2-k}{k+1} - \mu^{k/(1+k)}B^{-1} - \mu^{k/(1+k)} k gB^{-1}\approx 1  - \mu^{k/(1+k)}B^{-1} + \mu\frac{kL-k^2-k}{k+1}- \mu \frac{k(L-k-2)}{k+1}\\
&= 1 - \binom{L-1}{k}^{1/(k+1)}\mu^{k/(k+1)}+ \mu \frac{k}{k+1}.
\addtag
\end{align*}
If $L \gg k$, $m$ can be approximated as
\begin{align}
	m \approx 1 - (L\mu)^{k/(k+1)} (k!)^{-1/(k+1)} + \mu \frac{k}{k+1}.
\end{align}
\section{Mutational robustness on the mesa landscape in the absence of recombination}
Here we calculate the mutational robustness for the mesa landscape in the absence of
recombination and under the assumption that the mutation rate is
small. Here this is taken to imply that the genome-wide mutation rate $U
\equiv L\mu \ll 1$,
which implies that multiple mutations are negligible in the mutation step.
Using the same notation as before, the lumped equilibrium frequencies
after mutation $f_n$ and after selection $q_n$ then satisfy the relations
\begin{align}
	\bar w  = \sum_{n=0}^k f_n,\quad
	q_n = \frac{f_n}{\bar w},\quad
	f_n = ( 1 - U)q_n + U \frac{L-n+1}{L} q_{n-1} + U \frac{n+1}{L} q_{n+1},
	\label{Eq:EOM_mesa_nor}
\end{align}
where $q_n = 0$ for $n > k$ and $q_{-1} = 0$.
Since $f_n = 0$ for $n > k+1$, we have
\begin{align}
  \label{Eq:barw}
	\bar w = 1 - f_{k+1} = 1 - U \frac{L-k}{L} q_k.
\end{align}
This yields a closed set of equations for the $q_n$, which reads 
\begin{align}
	q_n \left [ 1 - U \left (1 - \frac{k}{L} \right ) q_k \right ] = ( 1 - U ) q_n + U \frac{L-n+1}{L} q_{n-1} + U \frac{n+1}{L} q_{n+1}
\end{align}
or
\begin{align}
\frac{n+1}{L} q_{n+1} = M_k q_n - \frac{L-n+1}{L} q_{n-1},
\label{Eq:q_rec}
\end{align}
with 
\begin{equation}
  \label{Eq:Mk}
	M_k = 1 - \frac{L-k}{L} q_k = \sum_{n=0}^{k-1} q_n + \frac{k}{L} q_k = 1-\left (1-\frac{k}{L} \right ) q_k.
\end{equation}
Note that $M_k$ can be interpreted as mutational robustness measured before mutation and 
after selection. 
Interestingly, $q_n$'s do not depend on $U$ if no multiple mutations are allowed.
Since mutational robustness after mutation is given by
\begin{align}
	\nonumber
m =& \sum_{n=0}^{k-1} f_n + \frac{k}{L} f_k = 1 - f_{k+1} - \frac{L-k}{L} f_k
	= \bar w \left ( 1 - \frac{L-k}{K} q_k \right )= M_k \bar w \\
	=& M_k - U M_k(1-M_k)= M_k ( 1 - U) + U M_k^2,
\label{Eq:mut_rob_nor}
\end{align}
it is sufficient to find $M_k$.

Defining $\xi_n \equiv (2L)^{n/2}  \binom{L}{n}^{-1} q_n/q_0$ and $y
\equiv M_k\sqrt{L/2}$, we obtain from \eqref{Eq:q_rec}
\begin{align}
	\left (1 - \frac{n}{L} \right ) \xi_{n+1} = 2 y \xi_n - 2 n \xi_{n-1}.
\label{Eq:x_rec}
\end{align}
We write down the first few terms for later purposes,
\begin{align}
	\xi_0 = 1,\quad \xi_1 = 2 y, \quad \xi_2 = (4 y^2 - 2) \frac{L}{ L - 1}.
\label{Eq:H_f}
\end{align}
If $n/L \ll 1$, Eq~\eqref{Eq:x_rec} is approximated as
\begin{align}
\xi_{n+1} = 2 y\xi_n - 2 n \xi_{n-1},
\end{align}
which is the recursion relation of the Hermite polynomials $H_n(y)$. Since $\xi_0 = H_0$ and
$\xi_1 = H_1$ for any $L$, we find the approximate solution for $\xi_n$ as $\xi_n = H_n(y)$
for $n \ll L$.  If $k/L \ll 1$, the Hermite polynomial becomes an 
accurate solution for all $n$.
Since $\xi_{k+1} = 0$ by definition and $\xi_n > 0$ for $n \le k$, $y$
should be the largest solution of the equation
\begin{align}
H_{k+1}(y) = 0.
\label{Eq:Her_zero}
\end{align}
% Since , $x$ should be the greatest zero of the $(k+1)$'th Hermite polynomial.
If we denote the largest zero of Eq~\eqref{Eq:Her_zero} by $
\sqrt{y_k/2}$, we thus conclude
\begin{align}
	M_k = \sqrt{\frac{y_k}{L}} + o(L^{-1/2}).
  \label{Eq:Mk_Hermite}
\end{align}
The first few zeros are given by  
\begin{align}
	y_1 = 1,\quad y_2 = 3,\quad y_3 = 3 + \sqrt{6},\quad 
	y_4 = 5 + \sqrt{10}. 
	\label{Eq:hk_ex}
\end{align}
The approximation can be compared to the exact solutions for $M_k$ which have been obtained up to $k=4$ by solving 
Eq~\eqref{Eq:EOM_mesa_nor}, 
\begin{align*}
	M_1 &= \frac{1}{\sqrt{L}},\quad
	M_2 = \frac{\sqrt{3L-2}}{L} = \sqrt{\frac{3}{L}} + O(L^{-3/2}),\\
	M_3 &= \frac{\sqrt{3 L-4+\sqrt{6 L^2-3 L+16}}}{L} = \left ( \frac{3+\sqrt{6}}{L}\right )^{1/2}
	+ O(L^{-3/2}),\\ 
	M_4&=\frac{\sqrt{5 L-10+\sqrt{10 L^2-5 L+76}}}{L} = \left ( \frac{5 + \sqrt{10}}{L}\right )^{1/2}
	+ O(L^{-3/2}),
\end{align*}
which are indeed consistent with Eq~\eqref{Eq:Mk_Hermite}
and the first four $y_k$'s in Eq~\eqref{Eq:hk_ex}.
Using Eq~\eqref{Eq:mut_rob_nor} the robustness after mutation is then
given by 
\begin{align}
m \approx \sqrt{\frac{y_k}{L}}(1-U) + U \frac{y_k}{L}.
	\label{Eq:mut_rob_no_r}
\end{align}

Now we consider the case of large $k$. If we still assume $1 \ll k \ll L$,
the above approximation is valid. 
Since the asymptotic behavior of the largest zero of $H_n(x)$ is 
$\sim \sqrt{2n+1}$ \cite[p. 132]{ccappendix}, we find $y_k \sim 4k$, which gives
\begin{align}
m \approx 2 \sqrt{\frac{k}{L}} ( 1 - U).
	\label{Eq:Hermite_sol}
\end{align}

The approximation leading to Eq~\eqref{Eq:Mk_Hermite} is however not valid if $k/L$ remains finite as $L \rightarrow \infty$. 
To treat this problem, we may refer to previous work on the mesa landscape \cite{Wolff2009}
that makes use of a maximum principle for permutation-invariant
fitness landscapes \cite{Hermisson2002}. This principle states that
the stationary population mean fitness $\bar{w}$ is given by
\begin{align}
	\bar{w} 
	= \max_{x\in[0,1]} \left \{ \omega(x) - U \left [ 1 - 2 \sqrt{x(1-x)} \right ] \right \},
\end{align}
where $\omega(x) = \lim_{L \to \infty} w_{xL}$ is the limiting value
of the fitness of a genotype with $n=xL$ mutations. To account for the
fact that genotypes with more than $k$ mutation are lethal, the
fitness function has to be taken to be 
$\omega(x) = 1$ if $x \leq x_0 \equiv k/L$ and $\omega(x) = -\infty$ if $x>x_0$, which
is slighlty different from the setting of
Ref.~\cite{Wolff2009}. Nevertheless the result for the stationary
fitness is the same, 
\begin{align}
	\bar{w} = \begin{cases}
		1 - U \left [ 1 - 2 \sqrt{x_0(1-x_0)} \right ], & \text{ if } x_0 < 1/2,\\
		1, & \text{ if } x_0 \ge 1/2.
	\end{cases}
\end{align}
Combining Eqs~\eqref{Eq:barw} and \eqref{Eq:Mk} we see that $M_k = 1 -
U^{-1}(1-\bar{w})$, and therefore  
\begin{align}
	M_k = 
	\begin{cases}
		2 \sqrt{x_0(1-x_0)}, & \text{ if } x_0 < 1/2,\\
		1, & \text{ if } x_0 \ge 1/2.
	\end{cases}
\end{align}
Note that the leading behavior of $M_k$ for small $x_0$ is the same as
the Hermite polynomial solution
Eq~\eqref{Eq:Hermite_sol}.

\section{Recombination weight on the mesa landscape with uniform crossover}
In order to efficiently compute the recombination weight for uniform
crossover on the mesa landscape, one has to exploit the permutation
invariance of the landscape. In the following we denote the recombination weight $\lambda_{\sigma}$ of genotype $\sigma$ as $\lambda(L,a,k,r)$, since it is fully defined by the sequence length $L$, the mesa width $k$, the Hamming distance $a\equiv d_\sigma$ to the wild type and the recombination rate $r$. 
To start with we first note that the Hamming distances between an
offspring genotype $\sigma$ and its parent genotypes $\kappa$, $\tau$
also determine the Hamming distance between both parent genotypes
through the relation \cite{Boerlijst1996}
\begin{align}
	d(\sigma, \kappa)+d(\sigma, \tau)=d(\kappa, \tau). 
	\label{Eq:hdr}
\end{align}
For the following it is convenient to introduce the variables $i$ and $j$ which represent the Hamming distance $d(\sigma, \kappa)$ and $d(\sigma, \tau)$, respectively.
Eq~\eqref{Eq:hdr} is useful since the Hamming distance $i+j$ between the parent genotypes determines their number of possible distinct offspring genotypes through recombination. Hence the probability that the offspring genotype $\sigma$ is generated by two genotypes at distance $i$ and $j$ is given by
\begin{align}
	\frac{1}{2^{i+j}}r+\frac{1-r}{2}\left (\delta_{i0}+\delta_{j0} \right ),
\end{align}
where the second term includes the possibility of no recombination for which at least one of the parent genotypes needs to be the same as the offspring genotype, see also Eq~(6) of the main text. 
Next we consider the number of genotypes at Hamming distance $i$ and $j$ as well as their respective fitness. The number of potential parent genotypes at Hamming distance $i$ is given by $\binom{L}{i}$ which can be rewritten as 
\begin{align}
	\binom{L}{i}= \sum_{x=0}^{i}\ \binom{a}{x} \binom{L-a}{i-x}=\sum_{x=\max(0,i+a-L)}^{\min(i,a)}\ \binom{a}{x} \binom{L-a}{i-x}.
\end{align}
We make use of the fact that in order to create a genotype at distance
$i$, we can mutate $x$ out of $a$ 1-alleles and $i-x$ out of $L-a$
0-alleles from the offspring genotype for which the number of
arrangements is given by a binomial coefficient. Since the sum might contain zero terms we can restrict the summation range further. Through this expression it is possible to relate to each genotype its fitness which is given by
\begin{align}
	w(k,(a-x)+(i-x))=\theta(k-(a-x)-(i-x)),
\end{align}
where $(a-x)+(i-x)$ denotes the number of 1-alleles in the parent
genotype and $\theta$ is the Heaviside step function with $\theta(0)=1$. After
choosing a parent genotype at distance $i$ the remaining number of
suitable parent genotypes at Hamming distance $j$ is thus given by
\begin{align}
	\sum_{y=0}^{j}\ \binom{a-x}{y} \binom{L-a-(i-x)}{j-y}=\sum_{y=\max(0, j+a-L+i-x)}^{\min(j,a-x)}\ \binom{a-x}{y} \binom{L-a-(i-x)}{j-y},
\end{align}
with fitness
\begin{align}
	w(k,(a-y)+(j-y))=\theta(k-(a-y)-(j-y)).
\end{align}
Since the allele of at least one parent genotype needs to coincide with the allele of the offspring genotype, the number of 1-alleles that one can mutate is reduced by $x$. The same logic applies to the number of 0-alleles one can mutate, which is reduced by $i-x$.
Finally in order to compute the recombination weight we have to sum
over all possible combinations of distances $(i,j)$ which are
restricted due to Eq~\eqref{Eq:hdr} to be in the range $0 \leq i+j \leq L$. For efficient computation one should avoid double counting of ordered pairs $(i,j)$ and $(j,i)$ which yield the same contribution to the recombination weight. Combining these considerations leads to a more efficient expression for the recombination weight on the mesa landscape,
\begin{align}\label{Eq:RecMesa}
	\begin{aligned} 
		\lambda(L,k,a,r) = & \frac{1}{2^L} \sum_{i=0}^{\lfloor L/2 \rfloor}\sum_{j=i}^{L-i}\sum_{x=\max(0,i+a-L)}^{\min(i,a)}\binom{a}{x}\binom{L-x}{i-x}\theta(k+2x-a-i)  \times \\
		&   \sum_{y=\max(0, j+a-L+i-x)}^{\min(j,a-x)}\binom{a-x}{y}\binom{L+x-a-i}{j-y} \theta(k+2y-a-j) 
		   \left[\frac{r}{2^{i+j}}(2-\delta_{ij})+(1-r)\delta_{i0}\right],
	\end{aligned}
\end{align}
where $\lfloor z \rfloor$ stands for the greatest integer that is less than or
equal to $z$. As explained in the main text $\lambda(L,k,a,r)$ depends linearly on the recombination rate $r$.
We use Eq~\eqref{Eq:RecMesa} for numerical calculations.

      \begin{figure}
	\centering
	\includegraphics[width=0.5\linewidth]{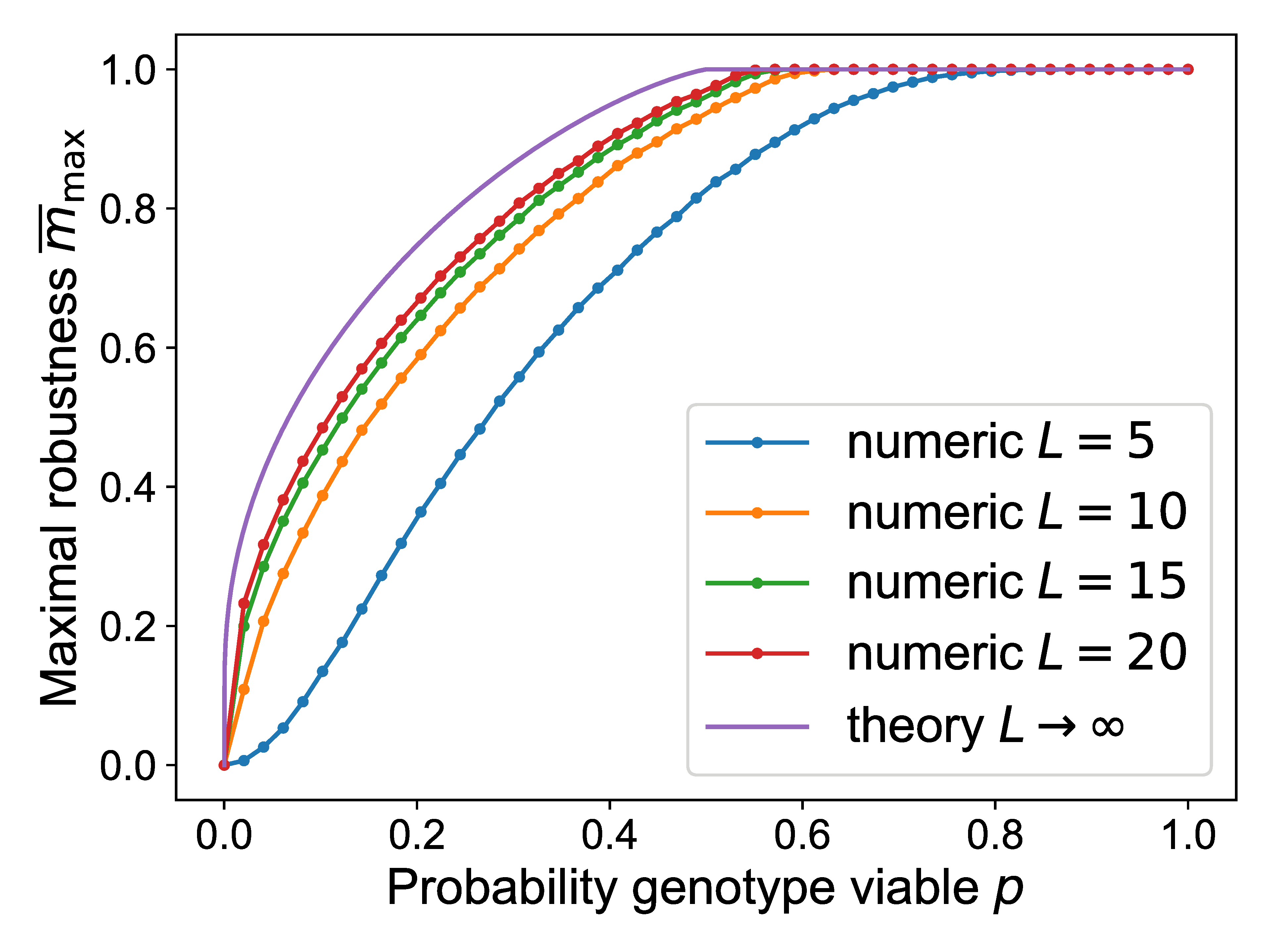}
	\caption[Source1]{\textbf{Maximal degree of the viable network
          in the percolation landscape.} The figure shows numerical
        results for the expected maximal degree of a viable genotype
        in percolation landscapes of different size $L$. For $L \to
        \infty$ the results converge to the solution $z^\ast$ of the
        equation $s_p(z^\ast) = \ln 2$, where $s_p(z)$ is given in
        Eq~\eqref{Eq:LD_bin}.} 
	\label{fig:maxdegree}
      \end{figure}
\section{Maximal robustness in the percolation landscape}

To estimate the number of viable neighbors of a genotype in the percolation
landscape in the limit of large $L$, we start from the observation
that the expected number of genotypes with $k$ viable neighbors is
\begin{equation}
	\mathbb{E}(n_k)  = 2^L \binom{L}{k} p^k (1-p)^{L-k} \sim \exp[L
  (\ln 2 - s_p(k/L))],
\end{equation}
where
\begin{equation}
  \label{Eq:LD_bin}
  s_p(z) = - z \ln(p) - (1-z) \ln(1-p) + z \ln(z) + (1-z) \ln(1-z)
\end{equation}
is the large deviation function of the binomial distribution
\cite{Sornette2000}. For a given $p$, there is thus a value $z^\ast(p)$
defined by $s_p(z^\ast) = \ln 2$ such that, for $L \to \infty$,
$\mathbb{E}(n_k) \to \infty$ if $k < z^\ast L$ and $\mathbb{E}(n_k)
\to 0$ if $k > z^\ast L$. Using standard probabilistic arguments 
this can be shown to imply that genotypes with $k$ neighbors are
present (absent) with probability 1 if $k < z^{\ast}L$ ($k > z^\ast
L$), respectively. Thus the expected maximal robustness is
$\overline{m}_{\mathrm{max}} = z^{\ast}$. Since $s_p(1)=\ln(1/p)$,
$z^{\ast} = 1$ for $p \geq \frac{1}{2}$. Fig~\ref{fig:maxdegree}
compares the asymptotic behavior of $\overline{m}_{\mathrm{max}}$ for $L
\to \infty$ to simulation results at finite $L$.
\bibliographystyle{apsrev}
\bibliography{preamble,mr}